\documentstyle[aps,amssymb]{revtex}

\begin{document}
\title{Glueball spectrum based on a rigorous three-dimensional relativistic
equation for two-gluon bound states I: Derivation of the relativistic
equation}
\author{Jun-Chen Su$^{1*,2}$ and Jian-Xing Chen$^3$}
\address{1. Department of Physics, Harbin Institute of technology, Harbin 150006,\\
People's Republic of China\\
2.Center for Theoretical Physics, College of Physics, Jilin\ University\\
Changchun 130023, People's Republic of China\\
3. Department of Physics, Peking University, Beijing 100871, People's\\
Republic of China}
\date{}
\maketitle

\begin{abstract}
A rigorous three-dimensional relativistic equation satisfied by two-gluon
bound states is derived from the QCD with massive gluons. With\ the gluon
fields and the quark fields being expanded in terms of the gluon multipole
fields and the spherical Dirac spinors respectively, the equation is well
established in the angular momentum representation and hence is\ much
convenient for solving the problem of two-gluon glueball spectra. In
particular, the interaction kernel in the equation is exactly derived and
given a closed expression which includes all the interactions taking place
in the two-gluon glueballs. The kernel contains only a few types of Green's
functions and commutators.\ Therefore, it is not only easily calculated by
the perturbation method, but also provides a suitable basis for
nonperturbative investigations.


*\ Corresponding author. E-mail address: junchens@public.cc.jl.cn
\end{abstract}

\section{Introduction}

It is a prominent feature of Quantum Chromodynamics (QCD) that besides the
quark-gluon interaction, there are also the interactions between gluons. The
self-interaction of gluons suggests that the gluons may form glueballs (the
bound states of gluons) through their interactions [1,2]. This is an
essential prediction which is of decisive significance for testing QCD and,
therefore, has been arising great interest in searching for the glueballs in
experiment [3-7]. But, there have been no faithful evidence to be
established so far for their existence [8-15]. On the other hand, the
property, the mass spectrum, the production and decay of the glueballs have
extensively been investigated theoretically. Many approaches were\ proposed
to serve such investigations such as the potential model [16-18], the bag
model [19-20], the sum rule [21], the Bethe-Salpeter (B-S) equation [22-24]
and the lattice simulation [25-30]. However, the theoretical results given
by different approaches are different and even contradictory with each other
[31-32]. This situation is attributed mainly to the fact that the
quark-gluon confinement has not clearly been understood so far. Commonly, it
is believed that the lattice gauge approach would give a more reliable
prediction because the approach is grounded on the first principle of QCD
and essentially nonperterbative although certain approximations are
inevitably made in practical calculations. In addition, it is widely
recognized that the B-S equation which is set up on the basis of quantum
field theory is a rigorous formalism for the bound state problem [33] and
suitable to study the glueballs [23-24]. Nevertheless, there are two
difficulties in the previous application of this equation. One difficulty
arises from the interaction kernel in the equation. This kernel was not
given a closed expression in the past although the expression can be derived
by the procedure as demonstrated in a recent publication of one author of
this paper [34]. Ordinarily, the kernel is defined by a sum of all
two-particle irreducible Feynman graphs and can only be calculated by the
perturbation method. Another difficulty is ascribed to the four-dimensional
nature of the equation in which the relative time (or the relative energy)
is unphysical and would lead to unphysical solutions [35]. So, many efforts
in the past are paid to recast the equation into a three-dimensional one in
the instantaneous approximation[36] or the quasipotential approach [37-38].

In this paper, we are devoted to deriving a rigorous three-dimensional
relativistic equation satisfied by two-gluon bound states so as to provide a
firm basis for further study. A practical application of this equation to
calculate the glueball spectrum will be presented in the next paper. The
distinctive features of the equation derived are as follows. (1) The
equation is exactly relativistic, containing all the retardation effect in
it, unlike the B-S equation given in the instantaneous approximation in
which the retardation effect is completely neglected. The equation is
derived in the equal-time formalism by the consideration that a bound state
is space-like\ and can exists in the equal-time Lorentz frame. In this
frame, the relativistic equation naturally becomes a three-dimensional one
without loss of any rigorism. Moreover, different from the B-S equation, the
three-dimensional equation derived in this paper is a standard
eigen-equation of Schr\"odinger-type. In the position space, it appears to
be first-order differential equations. In particular, the interaction kernel
in the equation is given a closed expression which is derived by the
procedure proposed first in Ref.[39] for a two-fermion system and
subsequently demonstrated in Ref.[40] for quark-antiquark bound states. The
kernel derived contains exactly all the interactions taking place in the
bound states and is represented in terms of only a few types of Green's
functions. Such a kernel can not only be easily calculated by the
perturbation method, but also is suitable for nonperturbative
investigations. (2) The three-dimensional equation\ is established in the
angular momentum representation. Since a glueball is the state of definite
spin and parity, obviously, to investigate the glueballs, it is much more
convenient to work in the angular momentum representation. In order to
express the relativistic equation in such a representation, it is necessary
to express the QCD in the same representation. This can be done by expanding
the quantized gluon and quark fields in terms of the gluon multipole fields
and the spherical Dirac spinors, respectively [41-42]. With these
expansions, the vertices in the interaction Hamiltonian can all be given
explicit and analytical expressions, as will be shown in the next paper.
This achievement is due to that the integrals containing three and four
spherical Bessel functions in the vertices are all calculated analytically
and expressed explicitly. We would like to note that in comparison with the
momentum representation in which every glueball state must be separately
constructed according to a certain requirement for the Lorentz and CTP
transformation properties [24] and the vertices would involve the gluon
polarization vectors which are not convenient to deal with, the angular
momentum representation has an advantage that the glueball state can easily
be written out in a consistent manner and the \ QCD vertices exhibit the
spin structures much clearly.(3) The relativistic equation is set up by
starting from the QCD with massive gluons. According to the conventional
concept of QCD, in order to keep the Lagrangian to be gauge-invariant, the
gluons must be massless. On the contrary, in most of the previous
investigations of the glueballs, an effective gluon mass was
phenomenologically introduced so as to get reasonable theoretical results
[17,18,23,24].\ The gluon mass was supposed to be generated dynamically from
the interaction with the physical vacuum of the Yang-Mills theory [23] or
through strong gluon-binding force [17]. Apparently, these arguments would
not be considered to be stringent and logically consistent with the concept
of the ordinary QCD. One of the authors of this paper in his recent article
[43] gave a different reasoning that the QCD with massive gluons can,
actually, be set up on the principle of gauge-invariance without the need of
introducing the Higgs mechanism or the St\"uckelberg fields. The essential
points to achieve this conclusion are: (a) The gluon fields \ must be viewed
as a constrained system in the whole space of vector potentials and the
Lorentz condition, as a necessary constraint, must be introduced from the
beginning and imposed on the Lagrangian; (b) The gauge-invariance\ of a
gauge field should be generally examined from the action of the field other
than from the Lagrangian because the action is of more fundamental dynamical
meaning than the Lagrangian. Particularly, for a constrained system such as
the gluon field, the gauge-invariance should be seen from its action given
in the physical space defined by the Lorentz condition. This concept is
well-known in Mechanics; (c) In the physical space, only infinitesimal gauge
transformations are possibly allowed and necessary to be considered. This
fact was clarified originally in Ref.[44]. Based on these points of view, it
is easy to prove that the QCD with massive gluons is gauge-invariant.
Moreover, the renormalizability and unitarity of the theory have been proved
to be no problems [45].

The remainder of this paper is arranged as follows. In Section II, the
massive QCD and its Lagrangian are briefly described. Section III is used to
formulate the angular momentum representation and give the expansions for
vector, spinor and ghost fields in this representation. In Section IV, the
expression of QCD Hamiltonian in the angular momentum space will be
described and discussed. Section V serves to derive the three-dimensional
relativistic equation satisfied by the glueballs. In Section VI, we are
devoted to derive a closed expression of the interaction kernel included in
the relativistic equation. In the last section, some remarks will be made.
In Appendix, we present a breif derivation of the spherical Dirac spinors
given in the angular momentum representation.

\section{QCD Lagrangian with massive gluons}

In the previous attempt of building up the massive non-Abelian gauge field
theory, the following massive Yang-Mills Lagrangian density was chosen to be
the starting point [43,46]: 
\begin{equation}
{\cal L}=-\frac 14F^{a\mu \nu }F_{\mu \nu }^a+\frac 12\mu ^2A^{a\mu }A_\mu
^a,  \eqnum{2.1}
\end{equation}
where $A_\mu ^a$ is the vector potential for a gluon field, 
\begin{equation}
F_{\mu \nu }^a=\partial _\mu A_\nu ^a-\partial _\nu A_\mu ^a+gf^{abc}A_\mu
^bA_\nu ^c  \eqnum{2.2}
\end{equation}
is the field strength tensor in which $g$ is the QCD coupling constant and $%
f^{abc}$ are the structure constants of color SU(3) group \ and $\mu $ is
the gluon mass. The first term in the Lagrangian is the ordinary Yang-Mills
Lagrangian which is gauge-invariant under the whole Lie group and used to
determine the form of interactions among the gluon fields themselves. The
second term in the Lagrangian is the mass term which is not gauge-invariant
and only affects the kinematic property of the fields. The above Lagrangian
itself was ever considered to give a complete description of the massive
gauge field dynamics. This consideration is not correct because the
Lagrangian is not only not gauge-invariant, but also contains redundant
unphysical degrees of freedom which must be eliminated by introducing a
suitable constraint condition. As we know, a massive gauge field has three
polarization states which need only three spatial components of the
four-dimensional vector potential $A_\mu ^a$ to describe them. In
Lorentz-covariant formulation, a full vector potential $A^\mu $ can be split
into two Lorentz-covariant parts: the transverse vector potential $A_T^\mu $
and the longitudinal vector potential $A_L^\mu $

\begin{equation}
A^\mu =A_T^\mu +A_L^\mu .  \eqnum{2.3}
\end{equation}
Since the Lorentz-covariant transverse vector potential $A_T^{a\mu }$
contains three-independent spatial components, it is sufficient to represent
the polarization states of a massive gluon. Whereas, the Lorentz-covariant
longitudinal vector potential $A_L^{a\mu }$ appears to be a redundant
unphysical variable which must be constrained by introducing the Lorentz
condition 
\begin{equation}
\partial ^\mu A_\mu ^a=0,  \eqnum{2.4}
\end{equation}
whose solution is 
\begin{equation}
A_L^{a\mu }=0.  \eqnum{2.5}
\end{equation}
With this solution, the massive Yang-Mills Lagrangian may be expressed in
terms of the independent dynamical variables $A_T^{a\mu }$%
\begin{equation}
{\cal L}=-\frac 14F_T^{a\mu \nu }F_{T\mu \nu }^a+\frac 12\mu ^2A_T^{a\mu
}A_{T\mu }^a,  \eqnum{2.6}
\end{equation}
which gives a complete description of the massive gluon field dynamics. If
we want to represent the dynamics in the whole space of the full vector
potential as described by the massive Yang-Mills Lagrangian in Eq.(2.1), the
massive gluon field must be treated as a constrained system. In this case,
the Lorentz condition in Eq.(2.4), as a constraint, is necessarily
introduced from the onset and imposed on the Lagrangian in Eq.(2.1) so as to
guarantee the redundant degrees of freedom to be eliminated from the
Lagrangian. Since the action is of more dynamical significance than the
Lagrangian, the gauge-invariance of QCD \ should generally be seen from the
action given by the Lagrangian in Eq.(2.6) or the Lagrangian in Eq.(2.1)
constrained by the Lorentz condition in Eq.(2.4). Under the gauge
transformations 
\begin{equation}
\delta A_\mu ^a=D_\mu ^{ab}\theta ^b,  \eqnum{2.7}
\end{equation}
where 
\begin{equation}
D_\mu ^{ab}=\delta ^{ab}\partial _\mu -gf^{abc}A_\mu ^c.  \eqnum{2.8}
\end{equation}
Noticing the identity $f^{abc}A^{a\mu }A_\mu ^b=0$, it \ is easy to prove
that the action given by the Lagrangian in Eq.(2.1) and constrained by he
Lorentz condition in Eq.(2.4) is gauge-invariant, 
\begin{equation}
\delta S=\int d^4x\delta {\cal L}=-\mu ^2\int d^4x\theta ^a\partial ^\mu
A_\mu ^a=0.  \eqnum{2.9}
\end{equation}
This suggests that the QCD with massive gluons may also be set up on the
basis of gauge-invariance principle.

Now, let us briefly describe quantization of the QCD with massive gluons,
This quantization was carried out by different approaches in Ref.[43]. A
simpler quantization is performed in the Lagrangian path-integral formalism
by means of the Lagrange undetermined multiplier method \ which was shown to
be equivalent to the Faddeev-Popov approach of quantization [44]. For this
quantization, it is convenient to generalize the\ QCD Lagrangian and the
Lorentz condition to the following forms: 
\begin{eqnarray}
{\cal L}_\lambda &=&\bar \psi \{i\gamma ^\mu (\partial _\mu -igT^aA_\mu
^a)-m\}\psi -\frac 14F^{a\mu \nu }F_{\mu \nu }^a  \nonumber \\
&&+\frac 12\mu ^2A^{a\mu }A_\mu ^a-\frac 12\alpha (\lambda ^a)^2 
\eqnum{2.10}
\end{eqnarray}
and

\begin{equation}
\partial ^\mu A_\mu ^a+\alpha \lambda ^a=0,  \eqnum{2.11}
\end{equation}
where for completeness, the quark fields have been included in the
Lagrangian in which $\bar \psi $ and $\psi $ stand for the quark fields, $%
T^a $ are the color matrices and $m$ is the quark mass, $\lambda ^a(x)$ are
the extra functions which will be identified with the Lagrange multipliers
and $\alpha $ is an arbitrary constant playing the role of gauge parameter.
According to the general procedure for constrained systems, \ the constraint
in Eq.(2.11) may be incorporated into the Lagrangian in Eq.(2.10) by the
Lagrange multiplier method, giving a generalized Lagrangian such that 
\begin{eqnarray}
{\cal L}_\lambda &=&\bar \psi \{i\gamma ^\mu (\partial _\mu -igT^aA_\mu
^a)-m\}\psi -\frac 14F^{a\mu \nu }F_{\mu \nu }^a+\frac 12\mu ^2A^{a\mu
}A_\mu ^a  \nonumber \\
&&\ -\frac 12\alpha (\lambda ^a)^2+\lambda ^a(\partial ^\mu A_\mu ^a+\alpha
\lambda ^a)  \nonumber \\
\ &=&\bar \psi \{i\gamma ^\mu (\partial _\mu -igT^aA_\mu ^a)-m\}\psi -\frac 1%
4F^{a\mu \nu }F_{\mu \nu }^a+\frac 12\mu ^2A^{a\mu }A_\mu ^a  \eqnum{2.12} \\
&&\ +\lambda ^a\partial ^\mu A_\mu ^a+\frac 12\alpha (\lambda ^a)^2. 
\nonumber
\end{eqnarray}
This Lagrangian is obviously not gauge-invariant. However, for building up a
correct gauge field theory, it is necessary to require the dynamics of the
system, i.e. the action given by the Lagrangian (2.12) to be invariant under
the gauge transformations\ denoted in Eqs.(2.7) and (2.8). By this
requirement, noticing the identity $f^{abc}A^{a\mu }A_\mu ^b=0$ and applying
the constraint condition in Eq.(2.11), we find

\begin{equation}
\delta S_\lambda =-\frac 1\alpha \int d^4x\partial ^\nu A_\nu ^a(x)\partial
^\mu ({\cal D}_\mu ^{ab}(x)\theta ^b(x))=0,  \eqnum{2.13}
\end{equation}
where 
\begin{equation}
{\cal D}_\mu ^{ab}(x)=\delta ^{ab}\frac{\sigma ^2}{\Box _x}\partial _\mu
^x+D_\mu ^{ab}(x),  \eqnum{2.14}
\end{equation}
in which $\sigma ^2=\alpha \mu ^2$ and $D_\mu ^{ab}(x)$ was defined in
Eq.(2.8). From equation (2.11) we see $\frac 1\alpha \partial ^\nu A_\nu
^a=-\lambda ^a\neq 0$. Therefore, to ensure the action to be
gauge-invariant, the following constraint condition on the gauge group is
necessary to be required 
\begin{equation}
\partial _x^\mu ({\cal D}_\mu ^{ab}(x)\theta ^b(x))=0.  \eqnum{2.15}
\end{equation}
These are the coupled equations satisfied by the parametric functions $%
\theta ^a(x)$ of the gauge group. Since the Jacobian is not singular 
\begin{equation}
detM\neq 0,  \eqnum{2.16}
\end{equation}
where 
\begin{eqnarray}
M^{ab}(x,y) &=&\frac{\delta (\partial _x^\mu {\cal D}_\mu ^{ac}(x)\theta
^c(x))}{\delta \theta ^b(y)}\mid _{\theta =0}  \nonumber \\
\  &=&\delta ^{ab}(\Box _x+\sigma ^2)\delta ^4(x-y)-gf^{abc}\partial _x^\mu
(A_\mu ^c(x)\delta ^4(x-y)),  \eqnum{2.17}
\end{eqnarray}
the above equations are solvable and would give a set of solutions which
express the parametric functions $\theta ^a(x)$ as functionals of the vector
potentials $A_\mu ^a(x)$. The constraint conditions in equation (2.15) may
also be incorporated into the Lagrangian (2.12) by the Lagrange undetermined
multiplier method. In doing this, it is convenient, as usually done, to
introduce ghost field variables $C^a(x)$ in such a fashion 
\begin{equation}
\theta ^a(x)=\xi C^a(x),  \eqnum{2.18}
\end{equation}
where $\xi $ is an infinitesimal Grassmann's number. In accordance with
Eq.(2.18), the constraint condition in Eq.(2.15) can be rewritten as 
\begin{equation}
\partial ^\mu ({\cal D}_\mu ^{ab}C^b)=0,  \eqnum{2.19}
\end{equation}
where the number $\xi $ has been dropped. This constraint condition usually
is called ghost equation. When the condition in Eq.(2.19) is incorporated
into the Lagrangian in Eq.(2.12) by the Lagrange multiplier method, we
obtain a more generalized Lagrangian as follows 
\begin{eqnarray}
{\cal L}_\lambda  &=&\bar \psi \{i\gamma ^\mu (\partial _\mu -igT^aA_\mu
^a)-m\}\psi -\frac 14F^{a\mu \nu }F_{\mu \nu }^a+\frac 12\mu ^2A^{a\mu
}A_\mu ^a  \nonumber \\
&&+\lambda ^a\partial ^\mu A_\mu ^a+\frac 12\alpha (\lambda ^a)^2+\bar C%
^a\partial ^\mu ({\cal D}_\mu ^{ab}C^b),  \eqnum{2.20}
\end{eqnarray}
where $\bar C^a(x)$, acting as Lagrange undetermined multipliers, are the
new scalar variables conjugate to the ghost variables $C^a(x).$

At present, we are ready to formulate the quantization of the QCD with
massive gluons. As we learn from the Lagrange undetermined multiplier
method, the dynamical and constrained variables as well as the Lagrange
multipliers in the Lagrangian (2.20) can all be treated as free ones,
varying arbitrarily. Therefore, we are allowed to use this kind of
Lagrangian to construct the generating functional of Green's functions 
\begin{eqnarray}
&&Z[J^{a\mu },\overline{\eta },\eta ,\overline{\xi }^a,\xi ^a]  \nonumber \\
&=&\frac 1N\int D(A_\mu ^a,\bar \psi ,\psi ,\bar C^a,C^a,\lambda
^a)exp\{i\int d^4x[{\cal L}_\lambda (x)+J^{a\mu }(x)A_\mu ^a(x)  \nonumber \\
&&\ +\ \bar \psi \eta +\overline{\eta }\psi +\overline{\xi }^a(x)C^a(x)+\bar 
C^a(x)\xi ^a(x)]\},  \eqnum{2.21}
\end{eqnarray}
where $D(A_\mu ^a,\cdots ,\lambda ^a)$ denotes the functional integration
measure, $J_\mu ^a,\overline{\eta },\eta ,\overline{\xi }^a$ and $\xi ^a$
are the external sources coupled to the gluon, quark and ghost fields and $N$
is a normalization constant. Looking at the expression of the Lagrangian
(2.20), we see, the integral over $\lambda ^a(x)$ is of Gaussian-type. Upon
completing the calculation of this integral, we finally arrive at [43] 
\begin{eqnarray}
&&Z[J^{a\mu },\overline{\eta },\eta ,\overline{\xi }^a,\xi ^a]  \nonumber \\
&=&\frac 1N\int D(A_\mu ^a,\bar \psi ,\psi ,\bar C^a,C^a,)exp\{i\int d^4x[%
{\cal L}_{eff}(x)  \nonumber \\
&&\ \ +J^{a\mu }(x)A_\mu ^a(x)+\bar \psi \eta +\overline{\eta }\psi +%
\overline{\xi }^a(x)C^a(x)+\bar C^a(x)\xi ^a(x)]\},  \eqnum{2.22}
\end{eqnarray}
where 
\begin{eqnarray}
{\cal L}_{eff} &=&\bar \psi \{i\gamma ^\mu (\partial _\mu -igT^aA_\mu
^a)-m\}\psi -\frac 14F^{a\mu \nu }F_{\mu \nu }^a+\frac 12\mu ^2A^{a\mu
}A_\mu ^a  \nonumber \\
&&-\frac 1{2\alpha }(\partial ^\mu A_\mu ^a)^2-\partial ^\mu \bar C^a{\cal D}%
_\mu ^{ab}C^b  \eqnum{2.23}
\end{eqnarray}
is the effective Lagrangian given in the general gauges.

From the generating functional shown in Eqs.(2.22) and ( 2.23), one may
derive the free gluon propagator as follows 
\begin{equation}
iD_{\mu \nu }^{cd}(k)=-\frac{i\delta ^{cd}}{k^2-\mu ^2+i\varepsilon }[g_{\mu
\nu }-(1-\alpha )\frac{k_\mu k_\nu }{k^2-\sigma ^2+i\varepsilon }]. 
\eqnum{2.24}
\end{equation}
It is emphasized that when the gluon mass tends to zero, this propagator
together with the effective Lagrangian in Eq.(2.23) and the generating
functional in Eq.(2.22) all immediately go over to the results given in the
QCD with massless gluons.

We would like to point out that the above propagator may also be derived
from the Lagrangian in Eq.(2.23) by the method of canonical
quantization[46]. In doing this, we need to use the Fourier representation
of the free gluon field operator[46] 
\begin{eqnarray}
{\bf A}_\mu ^c(x) &=&\int \frac{d^3k}{(2\pi )^{3/2}}\{\frac 1{2\omega (%
\overrightarrow{k})}\sum\limits_{\lambda =1}^3[\epsilon _\mu ^\lambda (%
\overrightarrow{k}){\bf a}_\lambda ^c(\overrightarrow{k})e^{-ikx}+\epsilon
_\mu ^{\lambda *}(\overrightarrow{k}){\bf a}_\lambda ^{c+}(\overrightarrow{k}%
)e^{ikx}]  \nonumber \\
&&\frac 1{2\omega _0(\overrightarrow{k})}\frac{\widetilde{k}}\mu [{\bf a}%
_0^c(\overrightarrow{k})e^{-i\widetilde{k}x}+{\bf a}_0^{c+}(\overrightarrow{k%
})e^{i\widetilde{k}x}],  \eqnum{2.25}
\end{eqnarray}
where $k=(k_0,\overrightarrow{k})$ and $\widetilde{k}=(\widetilde{k}_0,%
\overrightarrow{k})$ with $k_0\equiv \omega (\overrightarrow{k})=\sqrt{%
\overrightarrow{k}^2+\mu ^2}$ and $\widetilde{k}_0\equiv \omega _0(%
\overrightarrow{k})=\sqrt{\overrightarrow{k}^2+\sigma ^2}$, $\epsilon _\mu
^\lambda (\overrightarrow{k})$ are the polarization vectors satisfying the
transversity condition 
\begin{equation}
k^\mu \epsilon _\mu ^\lambda (\overrightarrow{k})=0,  \eqnum{2.26}
\end{equation}
which means that only three spatial components of $\epsilon _\mu ^\lambda (%
\overrightarrow{k})$ are independent and the creation and annihilation
operators ${\bf a}_\lambda ^c(\overrightarrow{k})$ and ${\bf a}_\lambda
^{c+}(\overrightarrow{k})$ ($\lambda =0,1,2,3)$ are subject to the following
commutation relations 
\begin{eqnarray}
\lbrack {\bf a}_\lambda ^c(\overrightarrow{k}),{\bf a}_{\lambda ^{\prime
}}^{d+}(\overrightarrow{k^{\prime }})] &=&-\delta ^{cd}g^{\lambda \lambda
^{\prime }}\delta ^3(\overrightarrow{k}-\overrightarrow{k^{\prime }}), 
\nonumber \\
\lbrack {\bf a}_\lambda ^c(\overrightarrow{k}),{\bf a}_{\lambda ^{\prime
}}^d(\overrightarrow{k^{\prime }})] &=&[{\bf a}_\lambda ^{c+}(%
\overrightarrow{k}),{\bf a}_{\lambda ^{\prime }}^{d+}(\overrightarrow{%
k^{\prime }})]=0.  \eqnum{2.27}
\end{eqnarray}
By making use of the expression in Eq.(2.25) and the above commutation
relations, it is not difficult to derive 
\begin{equation}
iD_{\mu \nu }^{cd}(x-y)=\left\langle 0\left| T\{{\bf A}_\mu ^c(x){\bf A}_\nu
^d(y)\}\right| 0\right\rangle =\int \frac{d^4k}{(2\pi )^4}iD_{\mu \nu
}^{cd}(k)e^{-ik(x-y)},  \eqnum{2.28}
\end{equation}
where $iD_{\mu \nu }^{cd}(k)$ is just as that written in Eq.(2.24).

The propagator in Eq.(2.24) is written in arbitrary gauges. It has been
proved that the S-matrices given by the QCD with massive gluons is
independent of the gauge parameter $\alpha $ [45].\ For example, in the tree
diagram approximation, noticing the transversity condition denoted in
Eq.(2.26) and the on-shell\ property of the gluon states, it is easy to
verify that\ the $\alpha $-dependent term \ proportional to $k_\mu k_\nu $
in Eq.(2.24) gives no contribution to the S-matrix elements. This fact was
early pointed out in Ref.[17]. In view of this fact, we may simply take the
Feynman gauge ($\alpha =1$) in practical calculations. In this gauge, the
effective Lagrangian in Eq.(2.23) becomes 
\begin{equation}
{\cal L}={\cal L}_0+{\cal L}_I,  \eqnum{2.29}
\end{equation}
where 
\begin{equation}
{\cal L}_0=\frac 12A_\nu ^a(\square +\mu ^2)A^{a\nu }+\bar \psi (i\gamma
^\mu \partial _\mu -m)\psi +\bar C^a(\square +\mu ^2)C^a  \eqnum{2.30}
\end{equation}
and 
\begin{eqnarray}
{\cal L}_I &=&-\frac 12gf^{abc}(\partial _\mu A_\nu ^a-\partial _\nu A_\mu
^a)A^{b\mu }A^{c\nu }-\frac 14g^2f^{abc}f^{ade}A^{b\mu }A^{c\nu }A_\mu
^dA_\nu ^e  \nonumber \\
&&+\bar \psi \gamma ^\mu T^a\psi A_\mu ^a+gf^{abc}\partial ^\mu \bar C%
^aC^bA_\mu ^c  \eqnum{2.31}
\end{eqnarray}
are the free and interaction parts of the Lagrangian respectively.
Correspondingly, the gluon propagator in Eq.(2.24) is reduced to 
\begin{equation}
iD_{\mu \nu }^{cd}(k)=-\frac{i\delta ^{cd}g_{\mu \nu }}{k^2-\mu
^2+i\varepsilon }.  \eqnum{2.32}
\end{equation}

\section{Expressions of the QCD fields in angular momentum representation}

The fields in the Lagrangian in Eqs.(2.30) and (2.31) may be expressed in
the angular momentum representation in terms of the eigenfunctions of total
angular momenta for the vector, spinor and scalar fields. These
eigenfunctions are described below. For the vector field, the complete set
of the eigenfunctions were already found in the literature[41,42]. They
include the scalar multipole field $A_{JM}^S(k\overrightarrow{x})$ and the
vectorial multipole fields $\overrightarrow{A}_{JM}^M(k\overrightarrow{x}),%
\overrightarrow{A}_{JM}^E(k\overrightarrow{x})$ and $\overrightarrow{A}%
_{JM}^L(k\overrightarrow{x}).$ They are displayed in the following. 
\begin{equation}
A_{JM}^S(k\overrightarrow{x})=\sqrt{\frac 2\pi }kj_J(kr)Y_{JM}(\widehat{x}),
\eqnum{3.1}
\end{equation}

\begin{eqnarray}
\overrightarrow{A}_{JM}^M(k\overrightarrow{x}) &=&\frac{-i}{\sqrt{J(J+1)}}%
\widehat{L}A_{JM}^S(k\overrightarrow{x})  \nonumber \\
&=&\sqrt{\frac 2\pi }kj_J(kr)\overrightarrow{Y}_{JJM}(\widehat{x}), 
\eqnum{3.2}
\end{eqnarray}
\begin{eqnarray}
\overrightarrow{A}_{JM}^E(k\overrightarrow{x}) &=&\frac 1{k\sqrt{J(J+1)}}%
\nabla \times \widehat{L}A_{JM}^S(k\overrightarrow{x})  \nonumber \\
&=&-i\sqrt{\frac 2\pi }\frac k{\sqrt{2J+1}}[\sqrt{J}j_{J+1}(kr)%
\overrightarrow{Y}_{JJ+1M}(\widehat{x})  \eqnum{3.3} \\
&&-\sqrt{J+1}j_{J-1}(kr)\overrightarrow{Y}_{JJ-1M}(\widehat{x})],  \nonumber
\end{eqnarray}

\begin{eqnarray}
\overrightarrow{A}_{JM}^L(k\overrightarrow{x}) &=&\frac{-i}k\nabla A_{JM}^S(k%
\overrightarrow{x})  \nonumber \\
&=&-i\sqrt{\frac 2\pi }\frac k{\sqrt{2J+1}}[\sqrt{J+1}j_{J+1}(kr)%
\overrightarrow{Y}_{JJ+1M}(\widehat{x})  \eqnum{3.4} \\
&&+\sqrt{J}j_{J-1}(kr)\overrightarrow{Y}_{JJ-1M}(\widehat{x})],  \nonumber
\end{eqnarray}
where $J,M$ mark the total angular momentum and its third component of the
vector field eigenfunction, $\overrightarrow{k}$ is the momentum of a single
particle, $k=|\overrightarrow{k}|,$ $j_l(kr)$ is the $l$-th spherical Bessel
function \ with $r=\left| \overrightarrow{x}\right| $, $\widehat{L}=-i%
\overrightarrow{x}\times \triangledown $ is the orbital angular momentum
operator and $\overrightarrow{Y}_{JlM}(\widehat{x})$ denotes the vectorial
spherical harmonic function \ with \ total angular momentum $JM$ and orbital
angular momentum \ $l=J-1,J,J+1.$ This function is defined as 
\begin{equation}
\overrightarrow{Y}_{JlM}(\widehat{x})=\sum\limits_{mq}C_{lm1q}^{JM}Y_{lm}(%
\widehat{x})\overrightarrow{e}_q,  \eqnum{3.5}
\end{equation}
\ where $C_{lm1q}^{JM}$ are the Clebsch-Gordan (C-G) coefficients, $Y_{lm}(%
\widehat{x})$ with $\widehat{x}=(\theta ,\phi )$ \ are the eigenfunction of
orbital angular momentum operator $\widehat{L}$ and $\overrightarrow{e}_q$
is the eigenfunction of the spin operator for a vector particle . Since the
function defined by 
\begin{equation}
\overrightarrow{A}_{JM}(\overrightarrow{x})=j_J(kr)\overrightarrow{Y}_{JlM}(%
\widehat{x})  \eqnum{3.6}
\end{equation}
and 
\begin{equation}
A_{JM}^0(\overrightarrow{x})=j_J(kr)Y_{JM}(\widehat{x})  \eqnum{3.7}
\end{equation}
respectively satisfy the following equations of motion derived from the
first term of the Lagrangian in Eq.(2.30) 
\begin{equation}
(\square +\mu ^2)\overrightarrow{A}=0  \eqnum{3.8}
\end{equation}
and 
\begin{equation}
(\square +\mu ^2)A^0=0  \eqnum{3.9}
\end{equation}
as we see, the functions in Eqs.(3.1)-(3.4) completely describe the
eigenstates of the total angular momentum for a free massive gluon field (or
for a massless gluon field when the mass $\mu =0)$. Furthermore, as the
functions $\overrightarrow{A}_{JM}^M(k\overrightarrow{x})$ and $%
\overrightarrow{A}_{JM}^E(k\overrightarrow{x})$ satisfy the transversity
condition 
\begin{equation}
\triangledown \cdot \overrightarrow{A}=0  \eqnum{3.10}
\end{equation}
and are related to each other by 
\begin{equation}
\nabla \times \overrightarrow{A}_{JM}^M(k\overrightarrow{x})=k%
\overrightarrow{A}_{JM}^E(k\overrightarrow{x}),  \eqnum{3.11}
\end{equation}
the $\overrightarrow{A}_{JM}^M(k\overrightarrow{x})$ is usually called
transverse magnetic multipole field and $\overrightarrow{A}_{JM}^E(k%
\overrightarrow{x})$ transverse electric multipole field. While the field $%
\overrightarrow{A}_{JM}^L(k\overrightarrow{x})$ is referred to as
longitudinal multipole field because it obeys the condition of longitudinal
fields $\triangledown \times \overrightarrow{A}=0.$ It is easy to verify
that these vectorial eigenfunctions meet the orthogonality relation 
\begin{equation}
\displaystyle \int 
d^3x\overrightarrow{A}_{JM}^\lambda (k\overrightarrow{x})\overrightarrow{A}%
_{J^{^{\prime }}M^{^{\prime }}}^{\lambda ^{^{\prime }}}(k^{^{\prime }}%
\overrightarrow{x})=\delta _{\lambda \lambda ^{^{\prime }}}\delta
(k-k^{^{\prime }})\delta _{JJ^{^{\prime }}}\delta _{MM^{^{\prime }}}, 
\eqnum{3.12}
\end{equation}
where $\lambda ,\lambda ^{\prime }=M,E,L$ label the different modes of the
multipole fields. Similarly, for\ the scalar multipole field, we have

\begin{equation}
\displaystyle \int 
d^3xA_{JM}^s(k\overrightarrow{x})A_{J^{^{\prime }}M^{^{\prime }}}^s(k%
\overrightarrow{x})=\delta (k-k^{^{\prime }})\delta _{JJ^{^{\prime }}}\delta
_{MM^{^{\prime }}}.  \eqnum{3.13}
\end{equation}
Therefore, the multipole fields defined in Eqs.(3.1)-(3.4) may suitably be
chosen as the basis functions to establish the angular momentum
representation for the gluon fields. A gluon field operator ${\bf A}_\mu
^c(x)$ may be expanded as 
\begin{equation}
\overrightarrow{{\bf A}}^c(x)=\sum\limits_{\lambda JM}\int_0^\infty \frac{dk%
}{\sqrt{2\omega }}[{\bf a}_{JM}^{c\lambda }(k)\overrightarrow{A}%
_{JM}^\lambda (k\overrightarrow{x})e^{-i\omega t}+{\bf a}_{JM}^{c\lambda
^{+}}(k)\overrightarrow{A}_{JM}^{\lambda ^{*}}(k\overrightarrow{x}%
)e^{i\omega t}],  \eqnum{3.14}
\end{equation}
where $\lambda =M,E,L$, 
\begin{equation}
{\bf A}_0^c(x)=\sum\limits_{JM}\int_0^\infty \frac{dk}{\sqrt{2\omega }}[{\bf %
a}_{JM}^{cs}(k)A_{JM}^s(k\overrightarrow{x})e^{-i\omega t}+{\bf a}%
_{JM}^{cs^{+}}(k)A_{JM}^{s^{*}}(k\overrightarrow{x})e^{i\omega t}]. 
\eqnum{3.15}
\end{equation}
The creation and annihilation operators in the above expansions satisfy the
following commutation relations 
\begin{eqnarray}
\lbrack {\bf a}_{JM}^{c\lambda }(k),{\bf a}_{J^{\prime }M^{\prime
}}^{c^{\prime }\lambda ^{\prime +}}(k^{\prime })] &=&\delta (k-k^{\prime
})\delta _{cc^{\prime }}\delta _{\lambda \lambda ^{\prime }}\delta
_{JJ^{\prime }}\delta _{MM^{\prime }},  \nonumber \\
\lbrack {\bf a}_{JM}^{c\lambda }(k),{\bf a}_{J^{\prime }M^{\prime
}}^{c^{\prime }\lambda ^{\prime }}(k^{\prime })] &=&[{\bf a}_{JM}^{c\lambda
+}(k),{\bf a}_{J^{\prime }M^{\prime }}^{c^{\prime }\lambda ^{\prime
+}}(k^{\prime })]=0,  \eqnum{3.16}
\end{eqnarray}
\begin{eqnarray}
\lbrack {\bf a}_{JM}^{cs}(k),{\bf a}_{J^{\prime }M^{\prime }}^{c^{\prime
}s^{\prime +}}(k^{\prime })] &=&\delta (k-k^{\prime })\delta _{cc^{\prime
}}\delta _{JJ^{\prime }}\delta _{MM^{\prime \prime }},  \nonumber \\
\lbrack {\bf a}_{JM}^{cs}(k),{\bf a}_{J^{\prime }M^{\prime }}^{c^{\prime
}s^{\prime }}(k^{\prime })] &=&[{\bf a}_{JM}^{cs+}(k),{\bf a}_{J^{\prime
}M^{\prime }}^{c^{\prime }s^{\prime +}}(k^{\prime })]=0.  \eqnum{3.17}
\end{eqnarray}

From the transformation of a gluon field under the space inversion [46] 
\begin{equation}
P\overrightarrow{{\bf A}}^c(t,\overrightarrow{x})P^{-1}=-\overrightarrow{%
{\bf A}}^c(t,-\overrightarrow{x}),P{\bf A}_o^c(t,\overrightarrow{x})P^{-1}=%
{\bf A}_o^c(t,-\overrightarrow{x}),  \eqnum{3.18}
\end{equation}
where $P$ is the space inversion operator and the parity of the multipole
fields 
\begin{equation}
A_{JM}^S(-k\overrightarrow{x})=(-1)^JA_{JM}^S(k\overrightarrow{x}),%
\overrightarrow{A}_{JM}^\lambda (-k\overrightarrow{x})=(-1)^{J+\pi _\lambda }%
\overrightarrow{A}_{JM}^\lambda (k\overrightarrow{x}),  \eqnum{3.19}
\end{equation}
where $\pi _\lambda =0$ if $\lambda =M$ and $\pi _\lambda =1$ if $\lambda
=E,L,$ one may find from Eqs.(3.14) and (3.15) the parities of the
annihilation operators 
\begin{equation}
P{\bf a}_{JM}^{cs}(k)P^{-1}=(-1)^J{\bf a}_{JM}^{cs}(k),P{\bf a}%
_{JM}^{c\lambda }(k)P^{-1}=(-1)^{J+1+\pi _\lambda }{\bf a}_{JM}^{c\lambda
}(k)  \eqnum{3.20}
\end{equation}
and the same parity for the creation operators. In addition, from the charge
conjugation of the gluon field $CA_\mu ^c(t,\overrightarrow{x})C^{-1}=-A_\mu
^c(t,\overrightarrow{x})$, we obtain a minus C-parity \ for the operators $%
{\bf a}_{JM}^{c\lambda }(k)$ and ${\bf a}_{JM}^{c\lambda +}(k)$ 
\begin{equation}
C{\bf a}_{JM}^{c\lambda }(k)C^{-1}=-{\bf a}_{JM}^{c\lambda }(k),C{\bf a}%
_{JM}^{c\lambda +}(k)C^{-1}=-{\bf a}_{JM}^{c\lambda +}(k),  \eqnum{3.21}
\end{equation}
where $\lambda =S,M,E,L.$

Now let us turn to the angular momentum representation of spinor fields. As
shown in Appendix, this representation\ may be set up by means of the
spherical Dirac spinors. These spinors are shown in the following. 
\begin{equation}
u_{JM}^\sigma (p\overrightarrow{x})=\left( 
\begin{array}{c}
\sqrt{\frac{\varepsilon +m}{2\varepsilon }}u_J^\sigma (pr)\Omega
_{JM}^\sigma (\widehat{x}) \\ 
-\sqrt{\frac{\varepsilon -m}{2\varepsilon }}u_J^{-\sigma }(pr)\Omega
_{JM}^{-\sigma }(\widehat{x})
\end{array}
\right) ,  \eqnum{3.22}
\end{equation}
\begin{equation}
v_{JM}^\sigma (p\overrightarrow{x})=(-1)^{J+M+\sigma }\left( 
\begin{array}{c}
\sqrt{\frac{\varepsilon -m}{2E}}u_J^{-\sigma }(pr)\Omega _{J-M}^{-\sigma }(%
\widehat{x}) \\ 
\sqrt{\frac{\varepsilon +m}{2\varepsilon }}u_J^\sigma (pr)\Omega
_{J-M}^\sigma (\widehat{x})
\end{array}
\right) ,  \eqnum{3.23}
\end{equation}
where $JM$ denote the total angular momentum and its third component of a
free fermion, $\sigma =\pm 1$, $\varepsilon =\sqrt{p^2+m^2}$ is the energy
of the fermion in which $m$ is the mass and $\overrightarrow{p}$ is the
momentum, $p=|\overrightarrow{p}|$, $\Omega _{JM}^\sigma (\widehat{x})$ is
the spherical harmonic spinor defined as 
\begin{equation}
\Omega _{JM}^\sigma (\widehat{x})=\left( 
\begin{array}{c}
\sigma \sqrt{\frac{J+\sigma (M-\frac 12)+\frac 12}{2J-l+1}}Y_{J-\frac \sigma 
2,M-\frac 12}(\widehat{x}) \\ 
\sqrt{\frac{J-\sigma (M+\frac 12)+\frac 12}{2J-\sigma +1}}Y_{J-\frac \sigma 2%
,M+\frac 12}(\widehat{x})
\end{array}
\right)  \eqnum{3.24}
\end{equation}
and $u_J^\sigma (pr)$ is defined by

\begin{equation}
u_J^\sigma (pr)=i^{J-\frac \sigma 2}\sqrt{\frac 2\pi }pj_{J-\frac \sigma 2%
}(pr).  \eqnum{3.25}
\end{equation}
The spinors $u_{JM}^\sigma (p\overrightarrow{x})$ and $v_{JM}^\sigma (p%
\overrightarrow{x})$ are the eigenstates of total angular momentum for a
fermion. They respectively obey the free Dirac equations of positive energy
state and negative energy state. Based on the above expressions of the
spherical Dirac spinors, it is easy to prove the following orthonormality
and completeness relations 
\begin{equation}
\begin{array}{l}
\int d^3xu_{JM}^{\sigma ^{+}}(p\overrightarrow{x})u_{J^{\prime }M^{\prime
}}^\sigma (p^{\prime }\overrightarrow{x})=\delta (p-p^{\prime })\delta
_{JJ^{^{\prime }}}\delta _{\sigma \sigma ^{\prime }}\delta _{MM^{\prime }},
\\ 
\int d^3xv_{JM}^{\sigma ^{+}}(p\overrightarrow{x})v_{J^{\prime }M^{\prime
}}^{\sigma ^{\prime }}(p^{\prime }\overrightarrow{x})=\delta (p-p^{\prime
})\delta _{JJ^{^{\prime }}}\delta _{\sigma \sigma ^{\prime }}\delta
_{MM^{\prime }}, \\ 
\int d^3xu_{JM}^{\sigma ^{+}}(p\overrightarrow{x})v_{J^{\prime }M^{\prime
}}^{\sigma ^{\prime }}(p^{\prime }\overrightarrow{x})=\int
d^3xv_{JM}^{\sigma ^{+}}(p\overrightarrow{x})u_{J^{\prime }M^{\prime
}}^{\sigma ^{\prime }}(p^{\prime }\overrightarrow{x})=0, \\ 
\sum\limits_{JM\sigma }\int_0^\infty dp[u_{JM}^\sigma (p\overrightarrow{x}%
)u_{JM}^{\sigma ^{+}}(p\overrightarrow{x}^{\prime })+v_{JM}^\sigma (p%
\overrightarrow{x})v_{JM}^{\sigma ^{+}}(p\overrightarrow{x}^{\prime
})]=\delta ^3(\overrightarrow{x}-\overrightarrow{x}^{\prime }).
\end{array}
\eqnum{3.26}
\end{equation}
Clearly, the free quark field operators may be expanded in terms of the
spherical Dirac spinors 
\begin{equation}
{\bf \psi }(x)=\sum\limits_{s\sigma JM}\int_0^\infty dp[{\bf b}%
_{JM}^{s\sigma }(p)u_{JM}^\sigma (p\overrightarrow{x})e^{-i\varepsilon t}+%
{\bf d}_{JM}^{s\sigma ^{+}}(p)v_{JM}^\sigma (p\overrightarrow{x}%
)e^{i\varepsilon t}],  \eqnum{3.27}
\end{equation}
\begin{equation}
\overline{{\bf \psi }}(x)=\sum\limits_{s\sigma JM}\int_0^\infty dp[{\bf b}%
_{JM}^{s\sigma ^{+}}(p)\overline{u}_{JM}^\sigma (p\overrightarrow{x}%
)e^{i\varepsilon t}+{\bf d}_{JM}^{s\sigma }(p)\overline{v}_{JM}^\sigma (p%
\overrightarrow{x})e^{-i\varepsilon t}],  \eqnum{3.28}
\end{equation}
where $s=(c,f),$ $c$ and $f$ are the color and flavor indices for a quark,$%
\overline{u}_{JM}^\sigma (p\overrightarrow{x})=u_{JM}^\sigma (p%
\overrightarrow{x})^{+}\gamma _0,\overline{v}_{JM}^\sigma (p\overrightarrow{x%
})=v_{JM}^\sigma (p\overrightarrow{x})^{+}\gamma _0$ and ${\bf b}%
_{JM}^{s\sigma ^{+}}(p),{\bf d}_{JM}^{s\sigma ^{+}}(p)$ and ${\bf b}%
_{JM}^{s\sigma }(p),{\bf d}_{JM}^{s\sigma }(p)$ are the creation and
annihilation operators. These operators satisfy the following
anticommutation relations 
\begin{equation}
\begin{array}{l}
\{{\bf b}_{JM}^{s\sigma }(p),{\bf b}_{J^{\prime }M^{\prime }}^{s^{\prime
}\sigma ^{\prime }+}(p^{\prime })\}=\delta (p-p^{\prime })\delta
_{ss^{^{\prime }}}\delta _{\sigma \sigma ^{\prime }}\delta _{JJ^{\prime
}}\delta _{MM^{\prime }}, \\ 
\{{\bf d}_{JM}^{s\sigma }(p),{\bf d}_{J^{\prime }M^{\prime }}^{s^{\prime
}\sigma ^{\prime }+}(p^{\prime })\}=\delta (p-p^{\prime })\delta
_{ss^{^{\prime }}}\delta _{\sigma \sigma ^{\prime }}\delta _{JJ^{\prime
}}\delta _{MM^{\prime }},
\end{array}
\eqnum{3.29}
\end{equation}
with the other anticommutators being zero.

From the space inversion transformation of the quark field [46] 
\begin{equation}
P{\bf \psi }(\overrightarrow{x},t)P^{-1}=\eta _P\gamma ^0{\bf \psi }(-%
\overrightarrow{x},t)  \eqnum{3.30}
\end{equation}
and the relations 
\begin{eqnarray}
\gamma ^0u_{JM}^\sigma (-p\overrightarrow{x}) &=&(-1)^{J-\frac \sigma 2%
}u_{JM}^\sigma (p\overrightarrow{x}),\gamma ^0v_{JM}^\sigma (-p%
\overrightarrow{x})  \nonumber \\
&=&(-1)^{J-\frac \sigma 2+1}v_{JM}^\sigma (p\overrightarrow{x}), 
\eqnum{3.31}
\end{eqnarray}
it is easy to find from Eqs.(3.27) and (3.28) the parity of the quark
operators such that

\begin{equation}
\begin{array}{l}
P{\bf b}_{lm}^{s\lambda }(p)P^{-1}=\eta _P(-1)^{J-\frac \sigma 2}{\bf b}%
_{lm}^{s\lambda }(p), \\ 
P{\bf d}_{lm}^{s\lambda }(p)P^{-1}=\eta _P(-1)^{J+\frac \sigma 2+1}{\bf d}%
_{lm}^{s\lambda }(p), \\ 
P{\bf b}_{lm}^{s\lambda ^{+}}(p)P^{-1}=\eta _P(-1)^{J-\frac \sigma 2}{\bf b}%
_{lm}^{s\lambda ^{+}}(p), \\ 
P{\bf d}_{lm}^{s\lambda ^{+}}(p)P^{-1}=\eta _P(-1)^{J+\frac \sigma 2+1}{\bf d%
}_{lm}^{s\lambda ^{+}}(p),
\end{array}
\eqnum{3.32}
\end{equation}
where the phase factor $\eta _P$ \ usually is chosen to be $\eta _P$ $=1$.
By noticing the charge conjugation transformations [46] 
\begin{equation}
{\cal B}{\bf \psi }(x){\cal B}^{-1}=\eta _CC\overline{{\bf \psi }}(x), 
\eqnum{3.33}
\end{equation}
where $C=i\gamma ^2\gamma ^0$ and 
\begin{equation}
v_{JM}^\sigma (p\overrightarrow{x})=C\overline{u}_{JM}^\sigma (p%
\overrightarrow{x})^T,u_{JM}^\sigma (p\overrightarrow{x})=C\overline{v}%
_{JM}^\sigma (p\overrightarrow{x})^T,  \eqnum{3.34}
\end{equation}
the $C$-parities of the creation and annihilation operators are easily found
from Eqs.(3.27) and (3.28) 
\begin{equation}
\begin{array}{l}
{\cal B}b_{lm}^{s\lambda }(p){\cal B}^{-1}=\eta _Cd_{lm}^{s\lambda }(p),%
{\cal B}d_{lm}^{s\lambda }(p){}{\cal B}^{-1}=\eta _Cb_{lm}^{s\lambda }(p),
\\ 
{\cal B}b_{lm}^{s\lambda ^{+}}(p){\cal B}^{-1}=\eta _Cd_{lm}^{s\lambda
^{+}}(p),{\cal B}d_{lm}^{s\lambda ^{+}}(p){}{\cal B}^{-1}=\eta
_Cb_{lm}^{s\lambda ^{+}}(p).
\end{array}
\eqnum{3.35}
\end{equation}

For the ghost fields, considering their scalar character, we can write their
expressions in the angular momentum representation as follows 
\begin{equation}
{\bf C}^a(x)=\sum\limits_{JM}\int_0^\infty \frac{dk}{\sqrt{2\omega }}[{\bf c}%
_{JM}^a(k)A_{JM}^s(k\overrightarrow{x})e^{-i\omega t}+{\bf d}%
_{JM}^{a^{+}}(k)A_{JM}^{s^{*}}(k\overrightarrow{x})e^{i\omega t}], 
\eqnum{3.36}
\end{equation}

\begin{equation}
\overline{{\bf C}}^a(x)=\sum\limits_{JM}\int_0^\infty \frac{dk}{\sqrt{%
2\omega }}[{\bf d}_{JM}^a(k)A_{JM}^s(k\overrightarrow{x})e^{-i\omega t}+{\bf %
c}_{JM}^{a^{+}}(k)A_{JM}^{s^{*}}(k\overrightarrow{x})e^{i\omega t}]. 
\eqnum{3.37}
\end{equation}
The anticommutation relations for the ghost particle operators in the above
are 
\begin{eqnarray}
\{{\bf c}_{JM}^a(k),{\bf c}_{J^{\prime }M^{\prime }}^{a^{\prime
}+}(k^{\prime })\} &=&\delta _{aa^{\prime }}\delta _{JJ^{\prime }}\delta
_{MM^{\prime }}\delta (k-k^{\prime }),  \nonumber \\
\{{\bf d}_{JM}^a(k),{\bf d}_{J^{\prime }M^{\prime }}^{a^{\prime
}+}(k^{\prime })\} &=&\delta _{aa^{\prime }}\delta _{JJ^{\prime }}\delta
_{MM^{\prime }}\delta (k-k^{\prime }),  \eqnum{3.38}
\end{eqnarray}
with the other anticommutators being vanishing.

\section{QCD Hamiltonian in angular momentum representation}

In this section, we are devoted to discussing the QCD Hamiltonian in the
angular momentum representation. By virtue of the expansions given in
Eqs.(3.14), (3.15), (3.27), (3.28), (3.36) and (3.37), it is not difficult
to formulate the Hamiltonian derived from the Lorentz-covariant Lagrangian
written in Eqs.(2.30) and (2.31). However, due to presence of the field $%
A^0(x)$, the expression of the Hamiltonian is rather complicated.
Considering that the field $A^0(x)$ is unphysical and will eventually be
eliminated from the S-matrix and the B-S equation by the ghost fields as
ensured by the unitarity of the theory [45], for the sake of simplifying the
representation of Hamiltonian, we would rather to start with the Lagrangian
obtained from the Lagrangian in Eqs.(2.30) and (2.31) by \ setting the
unphysical fields $A^0(x),\overline{C}^a(x)$ and $C^a(x)$ to vanish, as was
commonly done in the lattice gauge calculations [25-30] and similarly done
in Ref.[23] where a massive QCD Lagrangian taken in the temporal gauge was
used for calculating the glueball spectrum in the framework of B-S equation.
The Lagrangian we start with is 
\begin{equation}
{\cal L}={\cal L}_0+{\cal L}_I,  \eqnum{4.1}
\end{equation}
where 
\begin{eqnarray}
{\cal L}_0 &=&-\frac 12A_i^a(\square +\mu ^2)A_i^a+\bar \psi (i\gamma ^\mu
\partial _\mu -m)\psi  \nonumber \\
&=&\frac 12(\stackrel{\cdot }{A_k^a})^2+\frac 12A_i^a(\nabla ^2-\mu ^2)A_i^a+%
\bar \psi (i\gamma ^\mu \partial _\mu -m)\psi  \eqnum{4.2}
\end{eqnarray}
and 
\begin{eqnarray}
{\cal L}_I &=&-\frac 12gf^{abc}(\partial _iA_j^a-\partial _jA_i^a)A_i^bA_j^c
\nonumber \\
&&-\frac 14g^2f^{abc}f^{ade}A_i^bA_j^cA_i^dA_j^e-\bar \psi \gamma _iT^a\psi
A_i^a.  \eqnum{4.3}
\end{eqnarray}
By making use of the canonical variables conjugate to $A_k^a$, $\Psi $ and $%
\overline{\Psi }$ \ which are defined by 
\begin{eqnarray}
\Pi _k^a &=&\frac{\partial {\cal L}}{\partial \stackrel{\cdot }{A_k^a}}%
=A_k^a,  \nonumber \\
\Pi _\psi &=&\frac{\partial {\cal L}}{\partial \stackrel{\cdot }{\psi }}=i%
\overline{\psi }\gamma ^0,  \nonumber \\
\Pi _{\overline{\psi }} &=&\frac{\partial {\cal L}}{\partial \stackrel{\cdot 
}{\overline{\psi }}}=0,  \eqnum{4.4}
\end{eqnarray}
we can write the Hamiltonian density as 
\begin{eqnarray}
{\cal H} &=&\Pi _k^a\stackrel{\cdot }{A_k^a}+\Pi _\psi \stackrel{\cdot }{%
\psi }-{\cal L}  \nonumber \\
&=&{\cal H}_0+{\cal H}_I,  \eqnum{4.5}
\end{eqnarray}
where 
\begin{equation}
{\cal H}_0=\frac 12(\stackrel{\cdot }{A_k^a})^2-\frac 12A_i^a(\nabla ^2-\mu
^2)A_i^a+\bar \psi (\overrightarrow{\alpha }\cdot \triangledown +\beta m)\psi
\eqnum{4.6}
\end{equation}
and 
\begin{eqnarray}
{\cal H}_I &=&-\frac g2f^{abc}(\overrightarrow{A^a}\times \overrightarrow{A^b%
})\cdot (\nabla \times \overrightarrow{A^c})  \nonumber \\
&&+\frac 18g^2f^{abe}f^{cde}(\overrightarrow{A^a}\times \overrightarrow{A^b}%
)\cdot (\overrightarrow{A^c}\times \overrightarrow{A^d})  \eqnum{4.7} \\
&&+g\overline{\psi }T^a\overrightarrow{\gamma }\cdot \overrightarrow{A^a}%
\psi .  \nonumber
\end{eqnarray}
The first two terms in Eq.(4.6) denote the energy\ of \ free gluon fields.
The second term \ gives the energy of free quark fields. In the interaction
Hamiltonian density (4.7), the first, second and third terms represent the
gluon three-line vertex, the gluon four-line vertex and quark-gluon vertex
respectively. It should be noted that the above Hamiltonian includes not
only the transverse field $\overrightarrow{A}_T$\ , but also the
longitudinal field\ \ $\overrightarrow{A}_L$ \ whose presence can be seen
from the constraint condition in Eq.(2.11) because in the case of $A^0=0$
and $\alpha =1$, the condition becomes $\nabla \cdot \overrightarrow{A_L^c}%
=-\lambda \neq 0.$This is different from the theory proposed in Ref.[23]
where only the transverse field exists.

Now let us derive the expression of the Hamiltonian given by the Hamiltonian
density shown above in the angular momentum representation. First, we derive
the expression of the free Hamiltonian. On substituting
Eqs.(3.14),(3.15),(3.27),(3.28) into Eq.(4.6), one may get 
\begin{eqnarray}
H_0 &=&\int d^3x{\cal H}_0  \nonumber \\
&=&\sum\limits_{c\lambda JM}\int_0^\infty dk\omega (\overrightarrow{k}){\bf a%
}_{JM}^{c\lambda ^{+}}(k){\bf a}_{JM}^{c\lambda }(k)  \eqnum{4.8} \\
&&+\sum\limits_{s\sigma lm}\int_0^\infty dp\varepsilon (\overrightarrow{p})[%
{\bf b}_{JM}^{s\sigma ^{+}}(p){\bf b}_{JM}^{s\sigma }(p)+{\bf d}%
_{JM}^{s\sigma ^{+}}(p){\bf d}_{JM}^{s\sigma }(p)].  \nonumber
\end{eqnarray}

In later derivations, it is convenient to introduce compact notations for
various indices so as to simplify the expressions of the operators and
formulas. For the gluon field, we define 
\begin{equation}
{\bf a}_{JM}^{c\lambda \xi }(k)=\left\{ 
\begin{array}{c}
{\bf a}_{JM}^{c\lambda }(k)\text{ \ if }\xi =1 \\ 
\text{ \ }{\bf a}_{JM}^{c\lambda ^{+}}(k)\text{ if }\xi =-1
\end{array}
\right. ,  \eqnum{4.9}
\end{equation}
\begin{equation}
\overrightarrow{A}_{JM}^{\lambda \xi }(k\overrightarrow{x})=\left\{ 
\begin{array}{c}
\overrightarrow{A}_{JM}^\lambda (k\overrightarrow{x})\text{ \ if }\xi =1 \\ 
\text{ \ }\overrightarrow{A}_{JM}^{\lambda ^{*}}(k\overrightarrow{x})\text{
\ if }\xi =-1
\end{array}
\right. ,  \eqnum{4.10}
\end{equation}
\ then, the commutators in Eqs.(3.16) and (3.17) can be unified to write as 
\begin{equation}
\lbrack {\bf a}_{JM}^{c\lambda \xi }(k),{\bf a}_{J^{\prime }M^{\prime
}}^{c^{\prime }\lambda ^{\prime }\xi ^{\prime }}(k^{\prime })]=\delta
(k-k^{\prime })\delta _{cc^{\prime }}\delta _{\lambda \lambda ^{\prime
}}\delta _{JJ^{\prime }}\delta _{MM^{\prime }}\sin [(\xi -\xi ^{\prime })%
\frac \pi 4].  \eqnum{4.11}
\end{equation}
Furthermore, \ we define $\alpha \equiv (c,\lambda ,J,M,k,\xi )$ in which
the $\xi $ will be written as $\xi _\alpha $ later on. In this notation, the
commutator in Eq.(4.11) will be represented as 
\begin{equation}
\lbrack {\bf a}_\alpha ,{\bf a}_\beta ]=\triangle _{\alpha \beta }, 
\eqnum{4.12}
\end{equation}
where 
\begin{equation}
\triangle _{\alpha \beta }=\ \delta (k-k^{\prime })\delta _{cc^{\prime
}}\delta _{\lambda \lambda ^{\prime }}\delta _{JJ^{\prime }}\delta
_{MM^{\prime }}\sin [(\xi -\xi ^{\prime })\frac \pi 4]  \eqnum{4.13}
\end{equation}
and the expansion in Eq.(3.14) becomes 
\begin{equation}
\overrightarrow{A^c}(x)=\sum\limits_\alpha \ {\bf a}_\alpha \ 
\overrightarrow{A}_\alpha (k\overrightarrow{x})e^{-i\xi _\alpha \omega
_\alpha t},  \eqnum{4.14}
\end{equation}
where 
\begin{equation}
\sum\limits_\alpha \equiv \sum\limits_{c\lambda lm}\int_0^\infty dk,\ \omega
_\alpha =\sqrt{k^2+\mu ^2},  \eqnum{4.15}
\end{equation}
\begin{equation}
\overrightarrow{A}_\alpha (k\overrightarrow{x})=\frac 1{\sqrt{2\omega
_\alpha }}\overrightarrow{A}_{JM}^{\lambda \xi }(k\overrightarrow{x}). 
\eqnum{4.16}
\end{equation}

For the quark field, with the definition 
\begin{eqnarray}
{\bf c}_{lm}^{s\sigma \eta }(p) &=&\left\{ 
\begin{array}{cc}
{\bf b}_{lm}^{s\sigma }(p) & if\text{ \ }\eta {=1} \\ 
{\bf d}_{lm}^{s\sigma ^{+}}(p) & if\text{\ }\eta {=-1}
\end{array}
\right. ,  \eqnum{4.17} \\
\ w_{lm}^{\sigma \eta }(p\overrightarrow{x}) &=&\left\{ 
\begin{array}{cc}
u_{lm}^\sigma (p\overrightarrow{x}) & if\text{ }\eta {=1} \\ 
v_{lm}^\sigma (p\overrightarrow{x}) & if\text{ }\eta {=-1}
\end{array}
\right. ,  \eqnum{4.18}
\end{eqnarray}
\ the anticommutators in Eq.(3.29) becomes 
\begin{equation}
\{{\bf c}_{JM}^{s\sigma \eta }(p),{\bf c}_{J^{\prime ^{\prime }}M^{\prime
}}^{s^{\prime }\sigma ^{\prime }\eta ^{\prime }+}(p^{\prime })\}=\delta
(p-p^{\prime })\delta _{ss^{^{\prime }}}\delta _{\sigma \sigma ^{^{\prime
}}}\delta _{JJ^{\prime ^{\prime }}}\delta _{MM^{\prime }}\delta _{\eta \eta
^{\prime }}.  \eqnum{4.19}
\end{equation}
When we set $\alpha =(s,\sigma ,l,m,p,\eta )$ in which the $\eta $ will be
written as $\eta _\alpha $ later on, the above anticommutator can be simply
written as 
\begin{equation}
\{{\bf c}_\alpha ,{\bf c}_\beta ^{+}\}=\delta _{\alpha \beta },  \eqnum{4.20}
\end{equation}
where 
\begin{equation}
\delta _{\alpha \beta }\equiv \delta (p-p^{\prime })\delta _{ss^{^{\prime
}}}\delta _{\sigma \sigma ^{\prime }}\delta _{JJ^{\prime ^{\prime }}}\delta
_{MM^{\prime }}\delta _{\eta \eta ^{\prime }},  \eqnum{4.21}
\end{equation}
and the expansions in Eqs.(3.27) and (3.28) will be represented by 
\begin{equation}
\begin{array}{l}
{\bf \psi }(x)=\sum\limits_\alpha {\bf c}_\alpha w_\alpha (p\overrightarrow{x%
})e^{-i\xi _\alpha \varepsilon _\alpha t}, \\ 
\overline{{\bf \psi }}(x)=\sum\limits_\alpha {\bf c}_\alpha ^{+}\overline{w}%
_\alpha (p\overrightarrow{x})e^{i\xi _\alpha \varepsilon _\alpha t},
\end{array}
\eqnum{4.22}
\end{equation}
where 
\begin{equation}
\sum\limits_\alpha \equiv \sum\limits_{s\lambda lm}\int_0^\infty dp,\ \
\varepsilon _\alpha =\sqrt{p^2+m^2}.  \eqnum{4.23}
\end{equation}

In view of the notations defined above, the free Hamiltonian will compactly
be rewritten as follows 
\begin{equation}
H^0=H_g^0+H_q^0,  \eqnum{4.24}
\end{equation}
\begin{equation}
H_g^0=\frac 12\sum\limits_{\alpha \beta }\omega _{\alpha \beta }:{\bf a}%
_\alpha {\bf a}_\beta :,  \eqnum{4.25}
\end{equation}
\begin{equation}
H_q^0=\sum\limits_\alpha \xi _\alpha \varepsilon _\alpha :{\bf c}_\alpha ^{+}%
{\bf c}_\alpha :,  \eqnum{4.26}
\end{equation}
where the symbol $::$ denotes the normal product and 
\begin{equation}
\omega _{\alpha \beta }\equiv \ \omega (k)\delta (k-k^{\prime })\delta
_{cc^{\prime }}\delta _{\lambda \lambda ^{\prime }}\delta _{JJ^{\prime
}}\delta _{MM^{\prime }}(1-\delta _{\xi \xi ^{\prime }}).  \eqnum{4.27}
\end{equation}

Next, we derive the expression of the interaction Hamiltonian in the angular
momentum representation. In doing this, It is helpful to use the
transformations for an operator ( a Hamiltonian, a field operator, a
creation or annihilation operator) among the Heisenberg picture, the
Schr\"odinger picture and the interaction picture 
\begin{equation}
{\bf F}_H(t)=e^{iHt}{\bf F}_se^{-iHt}=e^{iHt}e^{-H_0t}{\bf F}%
_I(t)e^{iH_0t}e^{-iHt},  \eqnum{4.28}
\end{equation}
where ${\bf F}_H(t),{\bf F}_s$ and ${\bf F}_I(t)$ stand for the operators
given in the the Heisenberg, Schr\"odinger and interaction pictures. From
Eq.(4.28), it is clearly seen that ${\bf F}_s=$ ${\bf F}_H(0)={\bf F}_I(0)$
which is independent of time. According to the relation in Eq.(4.28), we
only need to write out the interaction Hamiltonian in the Schr\"odinger
picture (or in the interaction picture). In this picture, on inserting the
expressions in Eqs.(4.14) and (4.22) into Eq.(4.7), we obtain

\begin{eqnarray}
H_I(0) &=&\int d^3x{\cal H}_I(\overrightarrow{x},0)  \nonumber \\
&=&\sum\limits_{\alpha \beta \gamma }A(\alpha \beta \gamma ):{\bf a}_\alpha 
{\bf a}_\beta {\bf a}_\gamma :  \nonumber \\
+\sum\limits_{\alpha \beta \gamma \delta }B(\alpha \beta \gamma \delta ) &:&%
{\bf a}_\alpha {\bf a}_\beta {\bf a}_\gamma {\bf a}_\delta :  \eqnum{4.29} \\
+\sum\limits_{\alpha \beta \gamma }C(\alpha \beta \gamma ) &:&{\bf c}_\alpha
^{+}{\bf c}_\beta {\bf a}_\gamma :,  \nonumber
\end{eqnarray}
where 
\begin{equation}
A(\alpha \beta \gamma )=-\frac g2f^{abc}\int d^3x(\overrightarrow{A_\alpha }%
\times \overrightarrow{A_\beta })\cdot (\bigtriangledown \times 
\overrightarrow{A_\gamma }),  \eqnum{4.30}
\end{equation}
\begin{equation}
B(\alpha \beta \gamma \delta )=\frac 18g^2f^{abe}f^{cde}\int d^3x(%
\overrightarrow{A_\alpha }\times \overrightarrow{A_\beta })\cdot (%
\overrightarrow{A_\gamma }\times \overrightarrow{A_\delta })  \eqnum{4.31}
\end{equation}
and 
\begin{equation}
C(\alpha \beta \gamma )=g\int d^3x\overline{w}_\alpha T^c\overrightarrow{%
\gamma }w_\beta \cdot \overrightarrow{A_\gamma }.  \eqnum{4.32}
\end{equation}
By making use of the expressions of $\overrightarrow{A_\alpha },\overline{w}%
_\alpha $ and $w_\beta $, Completing the integrals over $\overrightarrow{x}$
and applying the formulas of angular momentum couplings, one may derive
explicit expressions of the coefficients $A(\alpha \beta \gamma ),B(\alpha
\beta \gamma \delta )$ and $C(\alpha \beta \gamma )$. The expressions of $%
A(\alpha \beta \gamma )$ and $B(\alpha \beta \gamma \delta )$ will be given
in the next paper. As for the coefficient $C(\alpha \beta \gamma ),$its
expression will be presented later in other publication.

\section{Three-dimensional relativistic equation for two gluon bound states}

\ The aim of this section is to derive a rigorous three-dimensional
relativistic equation satisfied by two-gluon bound states. This equation may
be derived from the equation satisfied by the following gluon two-time
Green's functions defined in the Heisenberg picture. These Green's \
functions include

\begin{eqnarray}
G(\alpha ^{+}\beta ^{+};\gamma ^{-}\delta ^{-};t_1-t_2) &=&\langle 0^{+}|T\{%
{\bf a}_\alpha (t_1){\bf a}_\beta (t_1){\bf a}_\gamma ^{+}(t_2){\bf a}%
_\delta ^{+}(t_2)\}|0^{-}\rangle ,  \nonumber \\
G(\alpha ^{-}\beta ^{-};\gamma ^{+}\delta ^{+};t_1-t_2) &=&\langle 0^{+}|T\{%
{\bf a}_\alpha ^{+}(t_1){\bf a}_\beta ^{+}(t_1){\bf a}_\gamma (t_2){\bf a}%
_\delta (t_2)\}|0^{-}\rangle ,  \nonumber \\
G(\alpha ^{-}\beta ^{+};\gamma ^{-}\delta ^{+};t_1-t_2) &=&\langle 0^{+}|T\{%
{\bf a}_\alpha ^{+}(t_1){\bf a}_\beta (t_1){\bf a}_\gamma ^{+}(t_2){\bf a}%
_\delta (t_2)\}|0^{-}\rangle ,  \eqnum{5.1} \\
G(\alpha ^{+}\beta ^{-};\gamma ^{+}\delta ^{-};t_1-t_2) &=&\langle 0^{+}|T\{%
{\bf a}_\alpha (t_1){\bf a}_\beta ^{+}(t_1){\bf a}_\gamma (t_2){\bf a}%
_\delta ^{+}(t_2)\}|0^{-}\rangle ,  \nonumber
\end{eqnarray}
where $T$ symbolizes the time-ordered product and $|0^{\pm }\rangle $ denote
the physical vacuum states defined in Heisenberg picture. It should be noted
that the creation and annihilation operators \ in Eq.(5.1) are those defined
in Eq.(3.15). For later convenience, we\ will use the notation defined in
Eq.(4.9) for the gluon operators. \ With this notation, instead of the above
Green's functions, we may start, in a consistent way, from the following
Green's function defined by 
\begin{equation}
G(\alpha \beta ;\gamma \delta ;t_1-t_2)=\langle 0^{+}|T\{{\bf a}_\alpha (t_1)%
{\bf a}_\beta (t_1){\bf a}_\gamma (t_2){\bf a}_\delta (t_2)\}|0^{-}\rangle ,
\eqnum{5.2}
\end{equation}
where ${\bf a}_\alpha $ can be a creation operator or an annihilation one.

On differentiating Eq.(5.2) with respect to time $t_{1,}$ it is found that 
\begin{equation}
\begin{array}{l}
i\frac \partial {\partial t_1}G(\alpha \beta ;\gamma \delta
;t_1-t_2)=i\delta (t_1-t_2)S(\alpha \beta ,\gamma \delta ) \\ 
+\langle 0^{+}|T\{[(i\frac \partial {\partial t_1}{\bf a}_\alpha (t_1)){\bf a%
}_\beta (t_1) \\ 
+{\bf a}_\alpha (t_1)(i\frac \partial {\partial t_1}{\bf a}_\beta (t_1))]%
{\bf a}_\gamma (t_2){\bf a}_\delta (t_2)\}|0^{-}\rangle ,
\end{array}
\eqnum{5.3}
\end{equation}
where 
\begin{equation}
\begin{array}{l}
S(\alpha \beta ,\gamma \delta )=\langle 0^{+}|[{\bf a}_\alpha (t_1){\bf a}%
_\beta (t_1),{\bf a}_\gamma (t_2){\bf a}_\delta (t_2)]|0^{-}\rangle \\ 
=\langle 0^{+}|[{\bf a}_\alpha {\bf a}_\beta ,{\bf a}_\gamma {\bf a}_\delta
]|0^{-}\rangle
\end{array}
\eqnum{5.4}
\end{equation}
is time-independent owing to the time displacement shown in Eq.(4.28) and
the restriction of $\delta (t_1-t_2)$. According to the relation in
Eq.(4.28), \ we have 
\begin{equation}
i\frac \partial {\partial t}{\bf a}_\alpha (t)=e^{iHt}[{\bf a}_\alpha
,H]e^{-iHt}.  \eqnum{5.5}
\end{equation}
Since ${\bf a}_\alpha $ in the commutator is independent of time, we may use
the Hamiltonian given in Eqs.(4.24)-(4.26) and (4.29) and the commutation
relation in Eq.(4.12) to compute the commutator $[{\bf a}_\alpha ,H]$. When
the result is substituted in Eq.(5.5), we obtain

\begin{equation}
\begin{array}{l}
i\frac \partial {\partial t_1}{\bf a}_\alpha (t_1)=\sum\limits_{\rho \sigma
}\omega _{\rho \sigma }\Delta _{\alpha \sigma }{\bf a}_\rho
(t_1)+\sum\limits_{\rho \sigma \tau }f_1(\rho \sigma \tau )\Delta _{\alpha
\tau }:{\bf a}_\rho (t_1){\bf a}_\sigma (t_1): \\ 
+\sum\limits_{\rho \sigma \tau \lambda }f_2(\rho \sigma \tau \lambda )\Delta
_{\alpha \lambda }:{\bf a}_\rho (t_1){\bf a}_\sigma (t_1){\bf a}_\tau (t_1):
\\ 
+\sum\limits_{\rho \sigma \tau }f_3(\rho \sigma \tau )\Delta _{\alpha \tau }:%
{\bf c}_\rho ^{+}(t_1){\bf c}_\sigma (t_1):,
\end{array}
\eqnum{5.6}
\end{equation}
where $\Delta _{\alpha \sigma }$ was defined in Eq.(4.13) and the
coefficients $f_i$ are defined by 
\begin{equation}
\begin{array}{l}
f_1(\rho \sigma \tau )=A(\rho \sigma \tau )+A(\rho \tau \sigma )+A(\tau \rho
\sigma ), \\ 
f_2(\rho \sigma \tau \lambda )=B(\rho \sigma \tau \lambda )+B(\rho \sigma
\lambda \tau )+B(\rho \lambda \sigma \tau )+B(\lambda \rho \sigma \tau ), \\ 
f_3(\rho \sigma \tau )=C(\rho \sigma \tau ).
\end{array}
\eqnum{5.7}
\end{equation}
Clearly, the expression of $i\frac \partial {\partial t_1}{\bf a}_\beta
(t_1) $ may be written out from Eq.(5.6) \ by replacing $\alpha $ \ with $%
\beta $.\ Inserting this expression and that given in Eq.(5.6) into
Eq.(5.3), one may find an equation of motion obeyed by the Green's function
denoted in Eq.(5.2) such that 
\begin{equation}
\begin{array}{l}
i\frac \partial {\partial t_1}G(\alpha \beta ;\gamma \delta
;t_1-t_2)=i\delta (t_1-t_2)S(\alpha \beta ,\gamma \delta ) \\ 
+\sum\limits_{\rho \sigma \lambda }\omega _{\rho \sigma }\Delta _{\alpha
\beta ,\sigma \lambda }G(\rho \lambda ;\gamma \delta ;t_1-t_2) \\ 
+\sum\limits_{\rho \sigma \lambda }g_1(\alpha \beta ;\rho \sigma \lambda
)G_1(\rho \sigma \lambda ;\gamma \delta ;t_1-t_2) \\ 
+\sum\limits_{\rho \sigma \tau \lambda }g_2(\alpha \beta ;\rho \sigma \tau
\lambda )G_2(\rho \sigma \tau \lambda ;\gamma \delta ;t_1-t_2) \\ 
+\sum\limits_{\rho \sigma \lambda }g_3(\alpha \beta ;\rho \sigma \lambda
)G_3(\rho \sigma \lambda ;\gamma \delta ;t_1-t_2),
\end{array}
\eqnum{5.8}
\end{equation}
where 
\begin{equation}
\Delta _{\alpha \beta ,\sigma \lambda }=\Delta _{\alpha \sigma }\delta
_{\beta \lambda }+\Delta _{\beta \sigma }\delta _{\alpha \lambda }, 
\eqnum{5.9}
\end{equation}
\begin{equation}
\begin{array}{l}
g_1(\alpha \beta ;\rho \sigma \lambda )=\sum\limits_\tau f_1(\rho \sigma
\tau )\Delta _{\alpha \beta ,\tau \lambda }, \\ 
g_2(\alpha \beta ;\rho \sigma \tau \lambda )=\sum\limits_\theta f_2(\rho
\sigma \tau \theta )\Delta _{\alpha \beta ,\theta \lambda }, \\ 
g_3(\alpha \beta ;\rho \sigma \lambda )=\sum\limits_\tau f_3(\rho \sigma
\tau )\Delta _{\alpha \beta ,\tau \lambda },
\end{array}
\eqnum{5.10}
\end{equation}
and the Green functions $G_i$ are defined by 
\begin{eqnarray}
&&G_1(\rho \sigma \lambda ;\gamma \delta ;t_1-t_2)  \nonumber \\
&=&\langle 0^{+}|T\{:{\bf a}_\rho (t_1){\bf a}_\sigma (t_1):{\bf a}_\lambda
(t_1){\bf a}_\gamma (t_2){\bf a}_\delta (t_2)\}|0^{-}\rangle ,  \eqnum{5.11}
\end{eqnarray}
\begin{eqnarray}
&&G_2(\rho \sigma \tau \theta ;\gamma \delta ;t_1-t_2)  \nonumber \\
&=&\langle 0^{+}|T\{:{\bf a}_\rho (t_1){\bf a}_\sigma (t_1){\bf a}_\tau
(t_1):{\bf a}_\theta (t_1){\bf a}_\gamma (t_2){\bf a}_\delta
(t_2)\}|0^{-}\rangle ,  \eqnum{5.12}
\end{eqnarray}
\begin{eqnarray}
&&G_3(\rho \sigma \lambda ;\gamma \delta ;t_1-t_2)  \nonumber \\
&=&\langle 0^{+}|T\{:{\bf c}_\rho ^{+}(t_1){\bf c}_\sigma (t_1):{\bf a}%
_\lambda (t_1){\bf a}_\gamma (t_2){\bf a}_\delta (t_2)\}|0^{-}\rangle . 
\eqnum{5.13}
\end{eqnarray}

By the Fourier transformation 
\begin{equation}
G_i(\alpha ,...,\delta ;\omega )=\frac 1i\int\nolimits_{-\infty }^{+\infty
}dte^{i\omega t}G_i(\alpha ,...,\delta ;t),  \eqnum{5.14}
\end{equation}
the equation in Eq.(5.8) will be \ of the form in the energy representation 
\begin{equation}
\begin{array}{l}
\omega G(\alpha \beta ;\gamma \delta ;\omega )=S(\alpha \beta ,\gamma \delta
) \\ 
+\sum\limits_{\rho \sigma \lambda }\omega _{\rho \sigma }\Delta _{\alpha
\beta ,\sigma \lambda }G(\rho \lambda ;\gamma \delta ;\omega ) \\ 
+\sum\limits_{\rho \sigma \lambda }g_1(\alpha \beta ;\rho \sigma \lambda
)G_1(\rho \sigma \lambda ;\gamma \delta ;\omega ) \\ 
+\sum\limits_{\rho \sigma \tau \lambda }g_2(\alpha \beta ;\rho \sigma \tau
\lambda )G_2(\rho \sigma \tau \lambda ;\gamma \delta ;\omega ) \\ 
+\sum\limits_{\rho \sigma \lambda }g_3(\alpha \beta ;\rho \sigma \lambda
)G_3(\rho \sigma \lambda ;\gamma \delta ;\omega ).
\end{array}
\eqnum{5.15}
\end{equation}
Let us look at the second term on the right hand side (RHS) of the above
equation. From the definitions of $\omega _{\rho \sigma }$ and $\Delta
_{\alpha \beta ,\sigma \lambda }$ given in Eqs.(4.27), (5.9) and (4.13), we
see, only when $\xi _\rho \neq \xi _\sigma ,$ $\omega _{\rho \sigma }$ and $%
\Delta _{\rho \sigma }$ are nonvanishing. When we define 
\begin{equation}
\omega _\alpha =\left\{ 
\begin{array}{c}
\omega (k)\text{ \ }if\text{ }\xi _\alpha =+1 \\ 
-\omega (k)\text{ \ }if\text{ }\xi _\alpha =-1
\end{array}
\right. ,  \eqnum{5.16}
\end{equation}
the second term mentioned above gives \ 
\begin{equation}
\sum\limits_{\rho \sigma \lambda }\omega _{\rho \sigma }\Delta _{\alpha
\beta ,\sigma \lambda }G(\rho \lambda ;\gamma \delta ;\omega )=(\omega
_\alpha +\omega _\beta )G(\alpha \beta ;\gamma \delta ;\omega ). 
\eqnum{5.17}
\end{equation}
Thus, Eq.(5.15) can be written as 
\begin{equation}
\begin{array}{l}
(\omega -\omega _\alpha -\omega _\beta )G(\alpha \beta ;\gamma \delta
;\omega ) \\ 
=S(\alpha \beta ,\gamma \delta )+\sum\limits_{\rho \sigma \lambda
}g_1(\alpha \beta ;\rho \sigma \lambda )G_1(\rho \sigma \lambda ;\gamma
\delta ;\omega ) \\ 
+\sum\limits_{\rho \sigma \tau \lambda }g_2(\alpha \beta ;\rho \sigma \tau
\lambda )G_2(\rho \sigma \tau \lambda ;\gamma \delta ;\omega ) \\ 
+\sum\limits_{\rho \sigma \lambda }g_3(\alpha \beta ;\rho \sigma \lambda
)G_3(\rho \sigma \lambda ;\gamma \delta ;\omega ).
\end{array}
\eqnum{5.18}
\end{equation}

In order to obtain a closed equation,\ it is necessary (actually possible)
to introduce effective interaction kernels $\Lambda _i$ in such a fashion 
\begin{eqnarray}
&&\sum\limits_{\rho \sigma \lambda }g_1(\alpha \beta ;\rho \sigma \lambda
)G_1(\rho \sigma \lambda ;\gamma \delta ;\omega )  \nonumber \\
&=&\sum\limits_{\rho \sigma }\Lambda _1(\alpha \beta ;\rho \sigma ;\omega
)G(\rho \sigma ;\gamma \delta ;\omega ),  \eqnum{5.19}
\end{eqnarray}
\begin{eqnarray}
&&\sum\limits_{\rho \sigma \tau \lambda }g_2(\alpha \beta ;\rho \sigma \tau
\lambda )G_2(\rho \sigma \tau \lambda ;\gamma \delta ;\omega )  \nonumber \\
&=&\sum\limits_{\rho \sigma }\Lambda _2(\alpha \beta ;\rho \sigma ;\omega
)G(\rho \sigma ;\gamma \delta ;\omega ),  \eqnum{5.20}
\end{eqnarray}
\begin{eqnarray}
&&\sum\limits_{\rho \sigma \lambda }g_3(\alpha \beta ;\rho \sigma \lambda
)G_3(\rho \sigma \lambda ;\gamma \delta ;\omega )  \nonumber \\
&=&\sum\limits_{\rho \sigma }\Lambda _3(\alpha \beta ;\rho \sigma ;\omega
)G(\rho \sigma ;\gamma \delta ;\omega ).  \eqnum{5.21}
\end{eqnarray}
With the kernels defined above, Eq.(5.18) becomes 
\begin{eqnarray}
&&(\omega -\omega _\alpha -\omega _\beta )G(\alpha \beta ;\gamma \delta
;\omega )  \nonumber \\
&=&S(\alpha \beta ,\gamma \delta )+\sum\limits_{\rho \sigma }K(\alpha \beta
;\rho \sigma ;\omega )G(\rho \sigma ;\gamma \delta ;\omega ),  \eqnum{5.22}
\end{eqnarray}
this is just the equation satisfied by the Green's function $G(\alpha \beta
;\gamma \delta ;\omega )$ in which 
\begin{equation}
K(\alpha \beta ;\rho \sigma ;\omega )=\sum\limits_{i=1}^3\Lambda _i(\alpha
\beta ;\rho \sigma ;\omega ).  \eqnum{5.23}
\end{equation}
is the total interaction kernel.

The equation obeyed by glueball states may be derived from the above
equation with the aid of the Lehmann representation of the Green's function.
Suppose $\{|n\rangle \}$ form a complete set of glueball states, noticing
the transformation in Eq.(4.28)\ and $H|n\rangle =E_n|n\rangle $, \ we can
write 
\begin{equation}
\begin{array}{l}
G(\alpha \beta ;\gamma \delta ;t_1-t_2) \\ 
=\sum\limits_n\{\theta (t_1-t_2)\langle 0^{+}|{\bf a}_\alpha {\bf a}_\beta
|n\rangle \langle n|{\bf a}_\gamma {\bf a}_\delta |0^{-}\rangle
e^{-iE_n(t_1-t_2)} \\ 
+\theta (t_2-t_1)\langle 0^{+}|{\bf a}_\gamma {\bf a}_\delta |n\rangle
\langle n|{\bf a}_\alpha {\bf a}_\beta |0^{-}\rangle e^{-iE_n(t_2-t_1)}\}.
\end{array}
\eqnum{5.24}
\end{equation}
Substituting in Eq.(5.24) the representation of the step function 
\begin{equation}
\theta (t)=\frac i{2\pi }\int dq\frac{e^{-iqt}}{q+i\epsilon }  \eqnum{5.25}
\end{equation}
and then performing a Fourier transformation \ with respect to time, we can
get from Eq.(5.24) the Lehmann representation such that 
\begin{equation}
G(\alpha \beta ;\gamma \delta ;\omega )=\sum\limits_n[\frac{\chi _{\alpha
\beta }(n)\overline{\chi }_{\gamma \delta }(n)}{\omega -E_n+i\epsilon }-%
\frac{\chi _{\gamma \delta }(n)\overline{\chi }_{\alpha \beta }(n)}{\omega
+E_n-i\epsilon }],  \eqnum{5.26}
\end{equation}
where $E_n$ is the total energy of a glueball state and 
\begin{equation}
\chi _{\alpha \beta }(n)=\langle 0^{+}|{\bf a}_\alpha {\bf a}_\beta
|n\rangle \ ,\ \ \ \ \overline{\chi }_{\alpha \beta }(n)=\langle n|{\bf a}%
_\alpha {\bf a}_\beta |0^{-}\rangle  \eqnum{5.27}
\end{equation}
are the B-S amplitudes\ describing the two-gluon\ bound states. On replacing
the $G(\alpha \beta ;\gamma \delta ;\omega )$ in Eq.(5.22)\ with its Lehmann
\ representation, then multiplying the both sides of Eq.(5.22) with $\omega
-E_n$ and finally taking the limit $\omega \longrightarrow E_n$, we
eventually arrive at 
\begin{equation}
(E_n-\omega _\alpha -\omega _\beta )\chi _{\alpha \beta
}(n)=\sum\limits_{\rho \sigma }K(\alpha \beta ;\rho \sigma ;E_n)\chi _{\rho
\sigma }(n).  \eqnum{5.28}
\end{equation}
This is just the wanted three-dimensional relativistic equation for the
two-gluon bound states. It should be pointed out that the above equation
actually represents a set of coupled equations which may be written out by
setting $\xi _\alpha $ in the index $\alpha $ and \ $\xi _\beta $ in the
index $\beta $ to be $\pm 1.$Each of the equations is manifestly of
Schr\"odinger-type, i.e., a standard eigen-equation. In the next paper,
these coupled equations will be reduced to an equivalent equation satisfied
only by the B-S amplitude for which the two gluons are in the positive
energy states.

\section{Closed expression of the interaction kernel}

In this section, we are devoted to derive the \ effective interaction kernel
appearing in Eq.(5.28) and defined in Eq.(5.23), giving a closed expression
of it. For this derivation, we need \ equations of motion which describe the
evolution of the Green's function $G(\alpha \beta ;\gamma \delta ;t_1-t_2)$
and those shown in Eqs.(5.11)-(5.13) with time $t_2$. Taking the derivative
of the Green's function in Eq.(5.2) with respect to $t_2$, we have 
\begin{equation}
\begin{array}{l}
i\frac \partial {\partial t_2}G(\alpha \beta ;\gamma \delta
;t_1-t_2)=-i\delta (t_1-t_2)S(\alpha \beta ,\gamma \delta ) \\ 
+\langle 0^{+}|T\{{\bf a}_\beta (t_1){\bf a}_\beta (t_1)[(i\frac \partial {%
\partial t_2}{\bf a}_\gamma (t_2)){\bf a}_\delta (t_2)+{\bf a}_\gamma (t_2)i%
\frac \partial {\partial t_2}{\bf a}_\delta (t_2)]\}|0^{-}\rangle ,
\end{array}
\eqnum{6.1}
\end{equation}
where 
\begin{equation}
\begin{array}{l}
i\frac \partial {\partial t_2}{\bf a}_\gamma (t_2)=\sum\limits_{\rho \sigma
}\omega _{\rho \sigma }\Delta _{\lambda \sigma }{\bf a}_\rho (t_2) \\ 
+\sum\limits_{\rho \sigma \tau }f_1(\rho \sigma \tau )\Delta _{\gamma \tau }:%
{\bf a}_\rho (t_2){\bf a}_\sigma (t_2): \\ 
+\sum\limits_{\rho \sigma \tau \lambda }f_2(\rho \sigma \tau \lambda )\Delta
_{\gamma \lambda }:{\bf a}_\rho (t_2){\bf a}_\sigma (t_2){\bf a}_\tau (t_2):
\\ 
+\sum\limits_{\rho \sigma \tau }f_3(\rho \sigma \tau )\Delta _{\gamma \tau }:%
{\bf c}_\rho ^{+}(t_2){\bf c}_\sigma (t_2):,
\end{array}
\eqnum{6.2}
\end{equation}
here the coefficients $f_i$ are the same as represented in Eqs.(5.7). The
expression of $i\frac \partial {\partial t_2}{\bf a}_\delta (t_2)$ can
directly be written out from Eq.(6.2) by replacing $\gamma $ by $\delta .$
When this expression and the one given \ above are inserted into Eq.(6.1),
we are led to 
\begin{equation}
\begin{array}{l}
i\frac \partial {\partial t_2}G(\alpha \beta ;\gamma \delta
;t_1-t_2)=-i\delta (t_1-t_2)S(\alpha \beta ,\gamma \delta ) \\ 
+\sum\limits_{\rho \sigma \lambda }\omega _{\rho \sigma }\Delta _{\gamma
\delta ,\sigma \lambda }G(\alpha \beta ;\rho \lambda ;t_1-t_2) \\ 
+\sum\limits_{\rho \sigma \lambda }G_1(\alpha \beta ;\rho \sigma \lambda
;t_1-t_2)g_1(\rho \sigma \lambda ;\gamma \delta ) \\ 
+\sum\limits_{\rho \sigma \tau \lambda }G_2(\alpha \beta ;\rho \sigma \tau
\lambda ;t_1-t_2)g_2(\rho \sigma \tau \lambda ;\gamma \delta ) \\ 
+\sum\limits_{\rho \sigma \lambda }G_3(\alpha \beta ;\rho \sigma \lambda
;t_1-t_2)g_3(\rho \sigma \lambda ;\gamma \delta ),
\end{array}
\eqnum{6.3}
\end{equation}
where $S(\alpha \beta ,\gamma \delta )$ was defined in Eq.(5.4), the
coefficients $g_i$ are the same as those represented in Eq.(5.10) since the
following equality holds 
\begin{equation}
g_i(\rho ,...,\lambda ;\gamma \delta )=g_i(\gamma \delta ;\rho ,...,\lambda )
\eqnum{6.4}
\end{equation}
and the Green functions $G_i$ are defined as follows 
\begin{equation}
\begin{array}{l}
G_1(\alpha \beta ;\rho \sigma \lambda ;t_1-t_2) \\ 
=\langle 0^{+}|T\{{\bf a}_\alpha (t_1){\bf a}_\beta (t_1):{\bf a}_\rho (t_2)%
{\bf a}_\sigma (t_2):{\bf a}_\lambda (t_2)\}|0^{-}\rangle , \\ 
G_2(\alpha \beta ;\rho \sigma \tau \lambda ;t_1-t_2) \\ 
=\langle 0^{+}|T\{{\bf a}_\alpha (t_1){\bf a}_\beta (t_1):{\bf a}_\rho (t_2)%
{\bf a}_\sigma (t_2){\bf a}_\tau (t_2):{\bf a}_\lambda (t_2)\}|0^{-}\rangle ,
\\ 
G_3(\alpha \beta ;\rho \sigma \lambda ;t_1-t_2) \\ 
=\langle 0^{+}|T\{{\bf a}_\alpha (t_1){\bf a}_\beta (t_1):{\bf c}_\rho
^{+}(t_2){\bf c}_\sigma (t_2):{\bf a}_\lambda (t_2)\}|0^{-}\rangle .
\end{array}
\eqnum{6.5}
\end{equation}
By the Fourier transformation denoted in Eq.(5.14) and noticing Eq.(5.17),
we obtain 
\begin{equation}
\begin{array}{l}
(\omega +\omega _\gamma +\omega _\delta )G(\alpha \beta ;\gamma \delta
;\omega )=S(\alpha \beta ,\gamma \delta ) \\ 
-\sum\limits_{\rho \sigma \lambda }G_1(\alpha \beta ;\rho \sigma \lambda
;\omega )g_1(\rho \sigma \lambda ;\gamma \delta ) \\ 
-\sum\limits_{\rho \sigma \tau \lambda }G_2(\alpha \beta ;\rho \sigma \tau
\lambda ;\omega )g_2(\rho \sigma \tau \lambda ;\gamma \delta ) \\ 
-\sum\limits_{\rho \sigma \lambda }G_3(\alpha \beta ;\rho \sigma \lambda
;\omega )g_3(\rho \sigma \lambda ;\gamma \delta ).
\end{array}
\eqnum{6.6}
\end{equation}

According to the same procedure as formulated in Eqs.(6.1)-(6.6), one can
derive the equations of motion obeyed by the Green's functions denoted in
Eqs.(5.11)-(5.13). Because the same expressions of the differentials $i\frac 
\partial {\partial t_2}{\bf a}_\gamma (t_2)$ and $i\frac \partial {\partial
t_2}{\bf a}_\delta (t_2)$ are employed in all the derivations, the equations
of motion for those Green's functions may immediately be written down by
referencing the equation in Eq.(6.6). The equation for the Green's function
in Eq.(5.11) is 
\begin{equation}
\begin{array}{l}
(\omega +\omega _\gamma +\omega _\delta )G_1(\rho \sigma \lambda ;\gamma
\delta ;\omega )=S_1(\rho \sigma \lambda ,\gamma \delta ) \\ 
-\sum\limits_{\mu \nu \tau }G_{11}(\rho \sigma \lambda ;\mu \nu \tau ;\omega
){}g_1(\mu \nu \tau ;\gamma \delta ) \\ 
-\sum\limits_{\mu \nu \tau \kappa }G_{12}(\rho \sigma \lambda ;\mu \nu \tau
\kappa ;\omega )g_2(\mu \nu \tau \kappa ;\gamma \delta ) \\ 
-\sum\limits_{\mu \nu \tau }G_{13}(\rho \sigma \lambda ;\mu \nu \tau ;\omega
)g_3(\mu \nu \tau ;\gamma \delta ),
\end{array}
\eqnum{6.7}
\end{equation}
where 
\begin{equation}
S_1(\rho \sigma \lambda ,\gamma \delta )=\langle 0^{+}|[:{\bf a}_\rho {\bf a}%
_\sigma :{\bf a}_\lambda ,{\bf a}_\gamma {\bf a}_\delta ]|0^{-}\rangle , 
\eqnum{6.8}
\end{equation}
$G_{11}(\rho \sigma \lambda ;\mu \nu \tau ;\omega ),G_{12}(\rho \sigma
\lambda ;\mu \nu \tau \kappa ;\omega )$ and $G_{13}(\rho \sigma \lambda ;\mu
\nu \tau ;\omega )$ are the Fourier transforms of the following Green's
functions 
\begin{equation}
\begin{array}{l}
G_{11}(\rho \sigma \lambda ;\mu \nu \tau ;t_1-t_2) \\ 
=\langle 0^{+}|T\{:{\bf a}_\rho (t_1){\bf a}_\sigma (t_1):{\bf a}_\lambda
(t_1):{\bf a}_\mu (t_2){\bf a}_\nu (t_2):{\bf a}_\tau (t_2)\}|0^{-}\rangle ,
\\ 
G_{12}(\rho \sigma \lambda ;\mu \nu \tau \kappa ;t_1-t_2) \\ 
=\langle 0^{+}|T\{:{\bf a}_\rho (t_1){\bf a}_\sigma (t_1):{\bf a}_\lambda
(t_1):{\bf a}_\mu (t_2){\bf a}_\nu (t_2){\bf a}_\tau (t_2):{\bf a}_\kappa
(t_2)\}|0^{-}\rangle , \\ 
G_{13}(\rho \sigma \lambda ;\mu \nu \tau ;t_1-t_2) \\ 
=\langle 0^{+}|T\{:{\bf a}_\rho (t_1){\bf a}_\sigma (t_1):{\bf a}_\lambda
(t_1):{\bf c}_\mu ^{+}(t_2){\bf c}_\nu (t_2):{\bf a}_\tau
(t_2)\}|0^{-}\rangle .
\end{array}
\eqnum{6.9}
\end{equation}
For the the Green's function in Eq.(5.12), we can write 
\begin{equation}
\begin{array}{l}
(\omega +\omega _\gamma +\omega _\delta )G_2(\rho \sigma \tau \lambda
;\gamma \delta ;\omega )=S_2(\rho \sigma \tau \lambda ,\gamma \delta ) \\ 
-\sum\limits_{\mu \nu \theta }G_{21}(\rho \sigma \tau \lambda ;\mu \nu
\theta ;\omega )g_1(\mu \nu \theta ;\gamma \delta ) \\ 
-\sum\limits_{\mu \nu \theta \kappa }G_{22}(\rho \sigma \tau \lambda ;\mu
\nu \kappa \theta ;\omega )g_2(\mu \nu \kappa \theta ;\gamma \delta ) \\ 
-\sum\limits_{\mu \nu \theta }G_{23}(\rho \sigma \tau \lambda ;\mu \nu
\theta ;\omega )g_3(\mu \nu \theta ;\gamma \delta ),
\end{array}
\eqnum{6.10}
\end{equation}
where 
\begin{equation}
S_2(\rho \sigma \tau \lambda ,\gamma \delta )=\langle 0^{+}|[:{\bf a}_\rho 
{\bf a}_\sigma {\bf a}_\tau :{\bf a}_\lambda ,{\bf a}_\gamma {\bf a}_\delta
]|0^{-}\rangle  \eqnum{6.11}
\end{equation}
and the Green's functions on the RHS of Eq.(6.10) are defined as 
\begin{equation}
\begin{array}{l}
G_{21}(\rho \sigma \tau \lambda ;\mu \nu \theta ;t_1-t_2) \\ 
=\langle 0^{+}|T\{:{\bf a}_\rho (t_1){\bf a}_\sigma (t_1){\bf a}_\tau (t_1):%
{\bf a}_\lambda (t_1):{\bf a}_\mu (t_2){\bf a}_\nu (t_2):{\bf a}_\theta
(t_2)\}|0^{-}\rangle , \\ 
G_{22}(\rho \sigma \tau \lambda ;\mu \nu \kappa \theta ;t_1-t_2) \\ 
=\langle 0^{+}|T\{:{\bf a}_\rho (t_1){\bf a}_\sigma (t_1){\bf a}_\tau (t_1):%
{\bf a}_\lambda (t_1):{\bf a}_\mu (t_2){\bf a}_\nu (t_2){\bf a}_\kappa (t_2):%
{\bf a}_\theta (t_2)\}|0^{-}\rangle , \\ 
G_{23}(\rho \sigma \tau \lambda ;\mu \nu \theta ;t_1-t_2) \\ 
=\langle 0^{+}|T\{:{\bf a}_\rho (t_1){\bf a}_\sigma (t_1){\bf a}_\tau (t_1):%
{\bf a}_\lambda (t_1):{\bf c}_\mu ^{+}(t_2){\bf c}_\nu (t_2):{\bf a}_\theta
(t_2)\}|0^{-}\rangle .
\end{array}
\eqnum{6.12}
\end{equation}
For the Green's function in Eq.(5.13), we have the equation 
\begin{equation}
\begin{array}{l}
(\omega +\omega _\gamma +\omega _\delta )G_3(\rho \sigma \lambda ;\gamma
\delta ;\omega )=S_3(\rho \sigma \lambda ,\gamma \delta ) \\ 
-\sum\limits_{\mu \nu \tau }G_{31}(\rho \sigma \lambda ;\mu \nu \tau ;\omega
){}g_1(\mu \nu \tau ;\gamma \delta ) \\ 
-\sum\limits_{\mu \nu \tau \kappa }G_{32}(\rho \sigma \lambda ;\mu \nu \tau
\kappa ;\omega ){}g_2(\mu \nu \tau \kappa ;\gamma \delta ) \\ 
-\sum\limits_{\mu \nu \tau }G_{33}(\rho \sigma \lambda ;\mu \nu \tau ;\omega
)g_3(\mu \nu \tau ;\gamma \delta ),
\end{array}
\eqnum{6.13}
\end{equation}
where 
\begin{equation}
S_3(\rho \sigma \lambda ,\gamma \delta )=\langle 0^{+}|[:{\bf c}_\rho ^{+}%
{\bf c}_\sigma :{\bf a}_\lambda ,{\bf a}_\gamma {\bf a}_\delta ]|0^{-}\rangle
\eqnum{6.14}
\end{equation}
and the Green's functions on the RHS of Eq.(6.13) are defined by 
\begin{equation}
\begin{array}{l}
G_{31}(\rho \sigma \lambda ;\mu \nu \tau ;t_1-t_2) \\ 
=\langle 0^{+}|T\{:{\bf c}_\rho ^{+}(t_1){\bf c}_\sigma (t_1):a_\lambda
(t_1):a_\mu (t_2)a_\nu (t_2):a_\tau (t_2)\}|0^{-}\rangle , \\ 
G_{32}(\rho \sigma \lambda ;\mu \nu \tau \kappa ;t_1-t_2) \\ 
=\langle 0^{+}|T\{:{\bf c}_\rho ^{+}(t_1){\bf c}_\sigma (t_1):{\bf a}%
_\lambda (t_1):{\bf a}_\mu (t_2){\bf a}_\nu (t_2){\bf a}_\tau (t_2):{\bf a}%
_\kappa (t_2)\}|0^{-}\rangle , \\ 
G_{33}(\rho \sigma \lambda ;\mu \nu \tau ;t_1-t_2) \\ 
=\langle 0^{+}|T\{:{\bf c}_\rho ^{+}(t_1){\bf c}_\sigma (t_1):{\bf a}%
_\lambda (t_1):{\bf c}_\mu ^{+}(t_2){\bf c}_\nu (t_2):{\bf a}_\tau
(t_2)\}|0^{-}\rangle .
\end{array}
\eqnum{6.15}
\end{equation}

Now we are ready to derive the interaction kernels. Multiplying the both
sides of Eq.(5.19) with $(\omega +\omega _\gamma +\omega _\delta )$ and then
applying Eqs.(6.6) and (6.7), we have 
\begin{eqnarray}
&&\sum\limits_{\rho \sigma }\Lambda _1(\alpha \beta ;\rho \sigma ;\omega
)[S(\rho \sigma ,\gamma \delta )-\sum\limits_{\mu \nu \lambda }G_1(\rho
\sigma ;\mu \nu \lambda ;\omega )g_1(\mu \nu \lambda ;\gamma \delta ) 
\nonumber \\
&&-\sum\limits_{\mu \nu \tau \lambda }G_2(\rho \sigma ;\mu \nu \tau \lambda
;\omega )g_2(\mu \nu \tau \lambda ;\gamma \delta )  \nonumber \\
&&-\sum\limits_{\mu \nu \lambda }G_3(\rho \sigma ;\mu \nu \lambda ;\omega
)g_3(\mu \nu \lambda ;\gamma \delta )]  \nonumber \\
&=&\sum\limits_{\rho \sigma \lambda }g_1(\alpha \beta ;\rho \sigma \lambda
)[S_1(\rho \sigma \lambda ,\gamma \delta )-\sum\limits_{\mu \nu \tau
}G_{11}(\rho \sigma \lambda ;\mu \nu \tau ;\omega )g_1(\mu \nu \tau ;\gamma
\delta )  \eqnum{6.16} \\
&&-\sum\limits_{\mu \nu \tau \kappa }G_{12}(\rho \sigma \lambda ;\mu \nu
\tau \kappa ;\omega )g_2(\mu \nu \tau \kappa ;\gamma \delta )  \nonumber \\
&&-\sum\limits_{\mu \nu \tau }G_{13}(\rho \sigma \lambda ;\mu \nu \tau
;\omega )g_3(\mu \nu \tau ;\gamma \delta )].  \nonumber
\end{eqnarray}

In order to obtain the expression of the kernel $\Lambda _1(\alpha \beta
;\rho \sigma ;\omega )$ from the above equation, it is necessary to
eliminate the kernel in the second, third and fourth terms on the left hand
side (LHS) of the above equation. Considering that the Green's function $%
G(\alpha \beta ;\gamma \delta ;\omega ),$as a matrix $G$, has an inverse $%
G^{-1}(\alpha \beta ;\gamma \delta ;\omega )$, then from Eq.(5.19) we can
get 
\begin{equation}
\Lambda _1(\alpha \beta ;\gamma \delta ;\omega )=\sum\limits_{\rho \sigma
\lambda }\sum\limits_{\mu \nu }g_1(\alpha \beta ;\rho \sigma \lambda
)G_1(\rho \sigma \lambda ;\mu \nu ;\omega )G^{-1}(\mu \nu ;\gamma \delta
;\omega ).  \eqnum{6.17}
\end{equation}
Substituting Eq.(6.17) into the second, third and fourth terms on the LHS of
Eq.(6.16), then moving these terms to the RHS of Eq.(6.16) and finally
acting on the both sides of that equation with the inverse of the matrix $%
S(\rho \sigma ,\gamma \delta )$ which is assumed to exist, \ we eventually
arrive at

\begin{eqnarray}
\Lambda _1(\alpha \beta ;\gamma \delta ;\omega ) &=&\Lambda _1^{(1)}(\alpha
\beta ;\gamma \delta ;\omega )+\Lambda _1^{(2)}(\alpha \beta ;\gamma \delta
;\omega )  \nonumber \\
&&+\Lambda _1^{(3)}(\alpha \beta ;\gamma \delta ;\omega ),  \eqnum{6.18}
\end{eqnarray}
where 
\begin{equation}
\Lambda _1^{(1)}(\alpha \beta ;\gamma \delta ;\omega )=\sum\limits_{\rho
\sigma \lambda }\sum\limits_{\mu \nu }g_1(\alpha \beta ;\rho \sigma \lambda
)S_1(\rho \sigma \lambda ,\mu \nu )S^{-1}(\mu \nu ,\gamma \delta ), 
\eqnum{6.19}
\end{equation}
\begin{eqnarray}
&&\Lambda _1^{(2)}(\alpha \beta ;\gamma \delta ;\omega )  \nonumber \\
&=&-\sum\limits_{\rho \sigma \lambda }\sum\limits_{\theta \pi }g_1(\alpha
\beta ;\rho \sigma \lambda )\{\sum\limits_{\mu \nu \tau }G_{11}(\rho \sigma
\lambda ;\mu \nu \tau ;\omega ){}g_1(\mu \nu \tau ;\theta \pi )  \nonumber \\
&&+\sum\limits_{\mu \nu \tau \kappa }G_{12}(\rho \sigma \lambda ;\mu \nu
\tau \kappa ;\omega )g_2(\mu \nu \tau \kappa ;\theta \pi )  \eqnum{6.20} \\
&&+\sum\limits_{\mu \nu \tau }G_{13}(\rho \sigma \lambda ;\mu \nu \tau
;\omega )g_3(\mu \nu \tau ;\theta \pi )\}S^{-1}(\theta \pi ,\gamma \delta ),
\nonumber
\end{eqnarray}
and 
\begin{eqnarray}
&&\Lambda _1^{(3)}(\alpha \beta ;\gamma \delta ;\omega )  \nonumber \\
&=&\sum\limits_{\rho \sigma \lambda }\sum\limits_{\mu \nu
}\sum\limits_{\epsilon \iota }\sum\limits_{\theta \pi }g_1(\alpha \beta
;\rho \sigma \lambda )G_1(\rho \sigma \lambda ;\mu \nu ;\omega )G^{-1}(\mu
\nu ;\epsilon \iota ;\omega )  \nonumber \\
&&\times \{\sum\limits_{\xi \eta \zeta }G_1(\epsilon \iota ;\xi \eta \zeta
;\omega )g_1(\xi \eta \zeta ;\theta \pi )  \nonumber \\
&&+\sum\limits_{\xi \eta \zeta \chi }G_2(\epsilon \iota ;\xi \eta \zeta \chi
;\omega )g_2(\xi \eta \zeta \chi ;\theta \pi )  \eqnum{6.21} \\
&&+\sum\limits_{\xi \eta \zeta }G_3(\epsilon \iota ;\xi \eta \zeta ;\omega
){}g_3(\xi \eta \zeta ;\theta \pi )\}S^{-1}(\theta \pi ,\gamma \delta ). 
\nonumber
\end{eqnarray}

By the same method as described above, the kernel $\Lambda _2(\alpha \beta
;\gamma \delta ;\omega )$ can be derived from the relation in Eq.(5.20) by
using the equations in Eq.(6.6) and (6.10). The result is\ as follows 
\begin{equation}
\Lambda _2(\alpha \beta ;\gamma \delta ;\omega )=\Lambda _2^{(1)}(\alpha
\beta ;\gamma \delta ;\omega )+\Lambda _2^{(2)}(\alpha \beta ;\gamma \delta
;\omega )+\Lambda _2^{(3)}(\alpha \beta ;\gamma \delta ;\omega ), 
\eqnum{6.22}
\end{equation}
where 
\begin{equation}
\Lambda _2^{(1)}(\alpha \beta ;\gamma \delta ;\omega )=\sum\limits_{\rho
\sigma \tau \lambda }\sum\limits_{\mu \nu }g_2(\alpha \beta ;\rho \sigma
\tau \lambda )S_2(\rho \sigma \tau \lambda ;\mu \nu ;\omega )S^{-1}(\mu \nu
,\gamma \delta ),  \eqnum{6.23}
\end{equation}
\begin{eqnarray}
&&\Lambda _2^{(2)}(\alpha \beta ;\gamma \delta ;\omega )  \nonumber \\
&=&-\sum\limits_{\rho \sigma \tau \lambda }\sum\limits_{\theta \pi
}g_2(\alpha \beta ;\rho \sigma \tau \lambda )\{\sum\limits_{\mu \nu
\varepsilon }G_{21}(\rho \sigma \tau \lambda ;\mu \nu \varepsilon ;\omega
){}g_1(\mu \nu \varepsilon ;\theta \pi )  \nonumber \\
&&+\sum\limits_{\mu \nu \varepsilon \kappa }G_{22}(\rho \sigma \tau \lambda
;\mu \nu \varepsilon \varkappa ;\omega )g_2(\mu \nu \varepsilon \varkappa
;\theta \pi )  \eqnum{6.24} \\
&&-\sum\limits_{\mu \nu \varepsilon }G_{23}(\rho \sigma \tau \lambda ;\mu
\nu \varepsilon ;\omega )g_3(\mu \nu \varepsilon ;\theta \pi
)\}S^{-1}(\theta \pi ,\gamma \delta ),  \nonumber
\end{eqnarray}
and 
\begin{eqnarray}
&&\Lambda _2^{(3)}(\alpha \beta ;\gamma \delta ;\omega )  \nonumber \\
&=&\sum\limits_{\rho \sigma \tau \lambda }\sum\limits_{\mu \nu
}\sum\limits_{\epsilon \iota }\sum\limits_{\theta \pi }g_2(\alpha \beta
;\rho \sigma \tau \lambda )G_2(\rho \sigma \tau \lambda ;\mu \nu ;\omega
)G^{-1}(\mu \nu ;\epsilon \iota ;\omega )  \nonumber \\
&&\times \{\sum\limits_{\xi \eta \zeta }G_1(\epsilon \iota ;\xi \eta \zeta
;\omega )g_1(\xi \eta \zeta ;\theta \pi )  \nonumber \\
&&+\sum\limits_{\xi \eta \zeta \kappa }G_2(\epsilon \iota ;\xi \eta \zeta
\chi ;\omega )g_2(\xi \eta \zeta \chi ;\theta \pi )  \eqnum{6.25} \\
&&+\sum\limits_{\xi \eta \zeta }G_3(\epsilon \iota ;\xi \eta \zeta ;\omega
)g_3(\xi \eta \zeta ;\theta \pi )\}S^{-1}(\theta \pi ,\gamma \delta ). 
\nonumber
\end{eqnarray}

Analogously, the kernel $\Lambda _3(\alpha \beta ;\gamma \delta ;\omega )$
may be derived from Eqs.(5.21), (6.6) and (6.13). The result is shown below 
\begin{eqnarray}
&&\Lambda _3(\alpha \beta ;\gamma \delta ;\omega )  \nonumber \\
&=&\Lambda _3^{(1)}(\alpha \beta ;\gamma \delta ;\omega )+\Lambda
_3^{(2)}(\alpha \beta ;\gamma \delta ;\omega )+\Lambda _3^{(3)}(\alpha \beta
;\gamma \delta ;\omega ),  \eqnum{6.26}
\end{eqnarray}
where 
\begin{equation}
\Lambda _3^{(1)}(\alpha \beta ;\gamma \delta ;\omega )=\sum\limits_{\rho
\sigma \lambda }\sum\limits_{\mu \nu }g_3(\alpha \beta ;\rho \sigma \lambda
)S_3(\rho \sigma \lambda ,\mu \nu )S^{-1}(\mu \nu ,\gamma \delta ), 
\eqnum{6.27}
\end{equation}
\begin{eqnarray}
&&\Lambda _3^{(2)}(\alpha \beta ;\gamma \delta ;\omega )  \nonumber \\
&=&-\sum\limits_{\rho \sigma \tau \lambda }\sum\limits_{\theta \pi
}g_3(\alpha \beta ;\rho \sigma \lambda )\{\sum\limits_{\mu \nu \tau
}G_{31}(\rho \sigma \lambda ;\mu \nu \tau ;\omega )g_1(\mu \nu \tau ;\theta
\pi )  \nonumber \\
&&+\sum\limits_{\mu \nu \tau \kappa }G_{32}(\rho \sigma \lambda ;\mu \nu
\tau \kappa ;\omega )g_2(\mu \nu \tau \kappa ;\theta \pi )  \eqnum{6.28} \\
&&+\sum\limits_{\mu \nu \tau }G_{33}(\rho \sigma \lambda ;\mu \nu \tau
;\omega )g_3(\mu \nu \tau ;\theta \pi )\}S^{-1}(\theta \pi ,\gamma \delta ),
\nonumber
\end{eqnarray}
\begin{eqnarray}
&&\Lambda _3^{(3)}(\alpha \beta ;\gamma \delta ;\omega )  \nonumber \\
&=&\sum\limits_{\rho \sigma \lambda }\sum\limits_{\mu \nu
}\sum\limits_{\epsilon \iota }\sum\limits_{\theta \pi }g_3(\alpha \beta
;\rho \sigma \lambda )G_3(\rho \sigma \lambda ;\mu \nu ;\omega )G^{-1}(\mu
\nu ;\epsilon \iota ;\omega )  \nonumber \\
&&\times \{\sum\limits_{\xi \eta \zeta }G_1(\epsilon \iota ;\xi \eta \zeta
;\omega )g_1(\xi \eta \zeta ;\theta \pi )  \nonumber \\
&&+\sum\limits_{\xi \eta \zeta \kappa }G_2(\epsilon \iota ;\xi \eta \zeta
\chi ;\omega )g_2(\xi \eta \zeta \chi ;\theta \pi )  \eqnum{6.29} \\
&&+\sum\limits_{\xi \eta \zeta }G_3(\epsilon \iota ;\xi \eta \zeta ;\omega
)g_3(\xi \eta \zeta ;\theta \pi )\}S^{-1}(\theta \pi ,\gamma \delta ). 
\nonumber
\end{eqnarray}

Based on the expressions described in Eqs.(6.18)-(6.29), the total kernel
can easily be written out. Using the matrix notation for the Green's
functions and the other functions, the kernel is represented as 
\begin{eqnarray}
K &=&\sum\limits_{i,j=1}^3\Lambda _i^{(i)}  \nonumber \\
\
&=&\{\sum\limits_{i=1}^3g_iS_i-\sum\limits_{i,j=1}^3g_iG_{ij}g_j+\sum%
\limits_{i,j=1}^3g_iG_iG^{-1}G_jg_j\}S^{-1}.  \eqnum{6.30}
\end{eqnarray}
This is just the closed expression of the interaction kernel. According to
the general argument as presented in Refs.[33,34,39,40], the third term in
the above expression plays the role of eliminating all the B-S reducible
(two-particle reducible) diagrams contained in the first two terms. In fact,
the relations in Eqs.(5.19)-(5.21) and the following ones 
\begin{equation}
G_jg_j=G\Lambda _j  \eqnum{6.31}
\end{equation}
are inserted \ into\ the last term in Eq.(6.30), it is seen that 
\begin{equation}
\sum\limits_{i,j=1}^3g_iG_iG^{-1}G_jg_j=\sum\limits_{i,j=1}^3\Lambda
_iG\Lambda _j=KGK,  \eqnum{6.32}
\end{equation}
which exhibits the typical structure of the B-S reducible part of the
interaction kernel. Therefore, the kernel shown in Eq.(6.30) is truly B-S
irreducible, consistent with the conventional concept. The equation in
Eq.(5.28) and the kernel in Eq.(6.30) will be employed, in the next paper,
to calculate the glueball spectrum in the ladder approximation.

\section{Concluding remarks}

In this paper, the exact three-dimensional relativistic equation for two
gluon glueball states and its interaction kernel have been derived from the
QCD with massive gluons. When the gluon mass tends to zero, the equation and
its kernel will naturally go over to the ones for the QCD with massless
gluons. As shown in Eq.(5.28), The equation derived is a standard eigenvalue
equation of Schr\"odinger-type. In the position space it appears to be a set
of first-order differential equations. This kind of equation was given for
fermion systems [39,40], but never formulated for boson systems in the past.
It should be noted that since the equation is derived in a special
equal-time Lorentz frame, it is certainly not Lorentz-covariant even though
the equation is rigorous and includes all the retardation effect in it. In
order to derive a Lorentz-covariant equation, one may start from the
four-time Green's function 
\begin{equation}
G_{\alpha \beta \gamma \delta }(T,t;T^{\prime },t^{\prime })=\langle
0^{+}|T\{{\bf a}_\alpha (t_1){\bf a}_\beta (t_2){\bf a}_\gamma (t_3){\bf a}%
_\delta (t_4)\}|0^{-}\rangle ,  \eqnum{7.1}
\end{equation}
where 
\begin{eqnarray}
T &=&\frac 12(t_1+t_2),t=t_1-t_2,  \nonumber \\
T^{\prime } &=&\frac 12(t_3+t_4),t^{\prime }=t_3-t_4.  \eqnum{7.2}
\end{eqnarray}
Differentiating the above Green's function with respect to $t_1$ and $t_2$
and employing the expression of the differential $i\frac \partial {\partial
t_1}{\bf a}_\alpha (t_1)$ as shown in Eq.(5.6) and the similar expression
for $i\frac \partial {\partial t_2}{\bf a}_\beta (t_2)$, one may obtain two
equations for the Green's function $G_{\alpha \beta \gamma \delta
}(T,t;T^{\prime },t^{\prime }).$ Adding the both equations\ and subtracting
one equation from another, following the same procedure as\ formulated in
Sec.V, it is easy to derive two equations satisfied by the B-S amplitude $%
\chi _{\alpha \beta }(T,t)$ which respectively describe the evolutions of
the state with respect to the center of mass time $T$ and the relative time $%
t$. In the equal-time frame, owing to the relative time being absent, we are
left only with the equation with respect to the center of mass time as given
in Sec.V. It is interesting to note that either the equation\ with respect
to $T$ or the equation with respect to $t$, appears to be a first order
differential equation of Schr\"odinger-type in the position space whose
solutions are determined \ merely\ by the initial condition of the B-S
amplitudes at the time origin. This is an essential feature of the equations
mentioned above which is different from the B-S equation. \ The latter
equation is a higher order differential equation and hence, like the
Klein-Gordon equation, has unphysical solutions with negative norm as
pointed out in the previous literature [35]. This is because the solutions
of B-S equation are determined not only by the initial amplitudes at the
time origin, but also by the time-differentials of the amplitudes at the
time origin.

Another point we would like to note is that unlike the Dyson-Schwinger
equation, the relativistic equation derived in this paper is of a closed
form. In particular, the interaction kernel in the equation is given a
closed \ expression. The expression contains only a few types of Green's
functions and vacuum expectation values\ of the operator commutators.\ They
are unambiguously defined in the Heisenberg picture and each of them can
independently be calculated by the perturbation method without concerning
other Green's functions. Especially, the kernel represents all the
interactions taking place in the bound states\ and, therefore, are suitable
for nonperturbative investigations because the Green's functions and vacuum
expectation values\ of the commutators, in principle, are able to be
evaluated by a certain nonperturbative method as suggested by the lattice
gauge approach.

At last, it would be pointed out that although the equation in Sec.V and the
kernel in Sec.VI are derived in the angular momentum representation, they
suit to formulate the equation and the kernel in the momentum representation
as long as the angular momentum quantum numbers in the indices $\alpha
,\beta ,\cdot \cdot \cdot $ are replaced by the momentum ones. That is to
say, the equation and the kernel formally remain \ unchanged in the both
representations.\ 

\section{Acknowledgement}

The authors are grateful to professor Shi-Shu Wu for useful discussions.
This work was supported in part by National Natural Science Foundation of
China.

\section{Appendix: Spherical spinors in the angular momentum representation}

In this appendix, we intend to give a derivation of the spherical Dirac
spinors which are used, as basis functions, to establish the angular
momentum representation for fermion fields. It is well-known that in the
relativistic case, unlike the helicity, the spin of a free fermion is not a
good quantum number. However, the total angular momentum operator of the
fermion commutes with the Hamiltonian. Therefore, it is meaningful to
discuss eigenfunctions of the total angular momentum which satisfy Dirac
equation. Let us start from the positive energy spinor $u_s(\overrightarrow{p%
})$ which is the\ solution to the Dirac equation $(i\partial ^\mu p_\mu -m)u(%
\overrightarrow{p})=0$. This spinor is taken to be [46] 
\begin{equation}
u_s(\overrightarrow{p})=\sqrt{\frac{\varepsilon +m}{2\varepsilon }}\left( 
\begin{tabular}{l}
$1$ \\ 
$\frac{\overrightarrow{\sigma }\cdot \overrightarrow{p}}{\varepsilon +m}$%
\end{tabular}
\right) \varphi _s  \eqnum{A.1}
\end{equation}
which is normalized in such a fashion: $u_s^{+}(\overrightarrow{p})u_s(%
\overrightarrow{p})=1$. The negative energy spinor can be given by the
charge conjugation $v_s(\overrightarrow{p})=C\overline{u}_s(\overrightarrow{p%
})^T$. Suppose $Y_{lm}(\widehat{p})$ with $\widehat{p}=\overrightarrow{p}%
/\left| \overrightarrow{p}\right| =(\theta ,\varphi )$ and $\varphi _s$ are
the orbital angular momentum and the spin eigenfunctions respectively; the
total angular momentum eigenfunctions may be constructed in the momentum
space by the C-G coupling 
\begin{equation}
\Omega _{JM}^l(\widehat{p})=\sum_{ls}C_{lm\frac 12s}^{JM}Y_{lm}(\widehat{p}%
)\varphi _s,  \eqnum{A.2}
\end{equation}
\ \ \ \ \ \ \ \ \ \ \ \ \ \ \ \ \ \ \ \ \ \ \ \ \ \ \ \ \ \ \ \ \ \ \ \ \ \
\ \ \ \ \ \ \ \ \ \ \ \ \ \ \ \ \ where $s=\frac \sigma 2,\sigma =\pm
1,l=J\pm \frac \sigma 2.$Noticing the representation 
\begin{equation}
\varphi _{\frac 12}=\left( 
\begin{tabular}{r}
0 \\ 
1
\end{tabular}
\right) ,\varphi _{-\frac 12}=\left( 
\begin{tabular}{r}
1 \\ 
0
\end{tabular}
\right)  \eqnum{A.3}
\end{equation}
and employing the explicit expressions of the C-G coupling coefficients for
different values of $\sigma $ which may be found in the textbook,\ one may
derive from (A.2) the expression\ as follows 
\begin{equation}
\Omega _{JM}^\sigma (\widehat{p})=\left( 
\begin{array}{c}
\sigma \sqrt{\frac{J+\sigma (M-\frac 12)+\frac 12}{2J-l+1}}Y_{J-\frac \sigma 
2,M-\frac 12}(\widehat{p}) \\ 
\sqrt{\frac{J-\sigma (M+\frac 12)+\frac 12}{2J-\sigma +1}}Y_{J-\frac \sigma 2%
,M+\frac 12}(\widehat{p})
\end{array}
\right) .  \eqnum{A.4}
\end{equation}

By making use of the \ Dirac spinor in (A.1) and the\ eigenfunctions
represented in (A.2) or in (A.4)\ with respect to the momentum $%
\overrightarrow{p}$, we may construct the\ spherical Dirac spinor in the \
position \ space through the following Fourier transformation 
\begin{equation}
u_{JM}^\sigma (p\overrightarrow{x})=\int d\widehat{p}\frac{e^{i%
\overrightarrow{p}\cdot \overrightarrow{x}}}{(2\pi )^{3/2}}pu(%
\overrightarrow{p})\Omega _{JM}^\sigma (\widehat{p}).  \eqnum{A.5}
\end{equation}
Substituting the expansion 
\begin{equation}
e^{i\overrightarrow{p}\cdot \overrightarrow{x}}=4\pi
\sum_{lm}i^lj_l(pr)Y_{lm}^{*}(\widehat{p})Y_{lm}(\widehat{x})  \eqnum{A.6}
\end{equation}
and the expressions \ written in (A.1) and (A.4) into (A.5), considering 
\begin{equation}
\overrightarrow{\sigma }\cdot \widehat{p}=\left( 
\begin{tabular}{ll}
$\cos \theta $ & $\sin \theta e^{-i\varphi }$ \\ 
$\sin \theta e^{i\varphi }$ & $-\cos \theta $%
\end{tabular}
\right)  \eqnum{A.7}
\end{equation}
and 
\begin{equation}
\overrightarrow{\sigma }\cdot \widehat{p}\Omega _{JM}^\sigma (\widehat{p}%
)=-\Omega _{JM}^{-\sigma }(\widehat{p}),  \eqnum{A.8}
\end{equation}
which is easily proved \ by utilizing the familiar recursion formulas for
the spherical harmornic functions, it is not difficult to derive the
expression shown in Eq.(3.22). The expression of the function $v_{JM}^\sigma
(p\overrightarrow{x})$ may be derived by the charge conjugation\ denoted in
Eq.(3.34). The result was written in Eq.(3.23). \ It would be noted that the
eigenfunction $\Omega _{JM}^\sigma (\widehat{x})$ in Eq.(3.24) which\ is
defined in the position space and appears in Eqs.(3.22) and (3.23) is of the
same form as the function $\Omega _{JM}^\sigma (\widehat{p})$ in (A.4) which
is defined in the momentum space.

\section{Reference}

[1] H.Fritzsch and M. Gell-Mann, in Proc.16 intern. Conf. on High Energy
Physics, FNAL, Batavia, IL, P.135 (1972).

[2] H. Fritzsch and P.Minkowski, Nuovo Cimento 30A, 393 (1975).

[3] Crystal Barrel Collaboration, Phys. Lett. B323, 223 (1994).

[4] V. V. Anisovich, et. al., Phys. Rev. D50, 1972 (1994).

[5] S. Bhatnagar and A. N. Mitra, Nuovo \ Cim. A104, 925 (1991).

[6] R. M. Baltrusaitis, et. al., Phys. Rev. Lett. 56, 107 (1986).

[7] J. Z. Bai, et. al., Phys. Rev. Lett. 76, 3502 (1996).

[8] R. M. Barnett, et. al., Phys. Rev. D54, 1 (1996).

[9] C. Amsler, et. al., Phys. Lett. B355, 425 (1995); Phys. Lett. B353, 385
(1995).

[10] D. Weingarten, Nucl. Phys. Proc. Suppl. B53, 232 (1997).

[11] D.V. Bugg, M. Peardon, B.S. Zou, Phys. Lett. B486, 49 (2000).

[12] J. Weinstein and N. Isgur, Phys. Rev. D27, 588 (1983); D41, 2236
(1990); D43, 95 (1991).

[13] L. Burakovsky, P.R. Page, Eur. Phys. J. C12, 489 (2000).

[14] J. Sexton, A. Vaccarino and D. Weigarten, Phys. Rev. Lett. 75, 4563
(1995).

[15] V. V. Anisovich, Phys. Lett. B364, 195 (1995).

[16] T. Barnes, Z. Phys. C10, 275 (1981).

[17] J. M. Cornwall and A. Soni, Phys. Lett. B120, 431 (1983).

[18] W. S. Hou and G. G. Wong, Phys. Rev. D67, 034003 (2003).

[19] R. L. Jaffe and K. Johnson, Phys. Lett. B60, 201 (1976); Phys. Rev.
Lett. 34, 1645 1976).

[20] T. Barnes, F. E. Close and S. Monagham, Nucl. Phys. B198, 380 (1982).

[21] S. Narision, Z. Phys. C26, 209 (1984);\ S. Narision, Nucl. Phys. B509,
312 1998).

[22] S. Bhatnagar and A. N. Mitra, Nuovo \ Cim. A104, 925 (1991).

[23] M. H. Thoma, M. L\"ust and H. J. Mang, J. Phys. G: Nucl. Part. Phys.
18, 1125 (1992).

[24] J. Y. Cui, J. M. Wu, H. Y. Jin, Phys. Lett. B424, 381 (1998).

[25] K.G. Wilson, Phys. Rev. D10, 2445 (1974).

[26] C. Bernard, Phys. Lett. 108B, 431 (1982); Nucl. Phys. B219, 341 (1983).

[27] C. Michael and M. Teper, Nucl. Phys. Lett. B314, 347 (1989).

[28] C. orningstar and M. Peardon, Phys. Rev. D60, 034509 (1999).

[29] C. Liu, Chin. Phys. Lett. 18, 187 (2001).

[30] D. Q. Liu, J. M. Wu, Y. Chen, High \ Ener. Phys. Nucl. Phys. 26, 222
(2002).

[31] G. B. West, talk given at the Montpelier Conference and Pairs Workshop
on QCD, hep-ph/9608258.

[32] W. Ochs, talk at EPS-HEP '99 Conference, July 15-21, 1999, Tampere,
Finland, hep-ph/9909241.

[33] E. E. Salpeter and H. A. Bethe, Phys. Rev. 84, 1232 (1951).

[34] J. C. Su, Commun. Theor. Phys. 38, 433 (2002).

[35] N. Nakanishi, Prog. Theor. Phys. Suppl. 42, 1 (1969); Prog. Theor.
Phys. Suppl. 95, 1 (1988).

[36] E. E. Salpeter, Phys. Rev. 87,328 (1952).

[37] R. Blankenbecler and R. Sugar, Phys. Rev. 142, 1051 (1969).

[38] I. T. Todorov, Phys. Rev. D3, 2351 (1971).

[39] S. S. Wu, J. Phys. G: Nucl. Part. Phys., 16, 1447 (1990).

[40] J. C. Su and D. Z. Mu, Commun. Theor. Phys. 15, 437 (1991).

[41] M. E. Rose, Multipole Fields, John Wiley \& Sons, New York, 1955.

[42] A. R. Edmonds, Angular \ Momentumin \ in \ Quantum Mechanics, Princeton
University Press, 1960.

[43] J. C. Su, Nuovo \ Cim. B 117, 203 (2002).

[44] L. D. Faddeev and V. N. Popov, Phys. Lett. B25, 29 (1967).

[45] J. C. Su, hep-th/9805193, hep-th/9805194.

[46] C.Itzykson and \ J-B. Zuber, Quantum \ Field Theory, McGraw-Hill, New
York, 1980.

\end{document}